\documentclass[11pt,a4paper]{article}
\usepackage[T1]{fontenc}
\usepackage[utf8]{inputenc}
\usepackage{lmodern}
\usepackage{amsmath,amssymb,amsthm}
\usepackage{graphicx}
\usepackage{booktabs}
\usepackage[margin=2.4cm]{geometry}
\usepackage{microtype}
\usepackage{placeins}

\newcommand{\E}{\mathbb{E}}
\newcommand{\Var}{\operatorname{Var}}
\newcommand{\Cov}{\operatorname{Cov}}
\newcommand{\Corr}{\operatorname{Corr}}
\newcommand{\Mult}{\operatorname{Mult}}
\newcommand{\sgn}{\operatorname{sgn}}

\newtheorem{theorem}{Theorem}[section]
\newtheorem{lemma}[theorem]{Lemma}
\newtheorem{proposition}[theorem]{Proposition}
\newtheorem{corollary}[theorem]{Corollary}
\newtheorem*{proposition*}{Proposition}
\theoremstyle{definition}
\newtheorem{remark}[theorem]{Remark}
\newtheorem{observation}[theorem]{Numerical observation}

\title{Properties of the discrete total variation distribution:\\
the limit law, its cumulants, and the tail approximation}
\author{
Nikola Bani\'c\\
Gideon Brothers, 10000 Zagreb, Croatia\\
\texttt{nbanic@gmail.com}
\and
Neven Elezovi\'c\\
Department of Applied Mathematics,\\
Faculty of Electrical Engineering and Computing,\\
University of Zagreb, 10000 Zagreb, Croatia\\
\texttt{neven.elezovic@fer.hr}
}
\date{}

\begin{document}
\maketitle

\begin{abstract}
\noindent
We study the null distribution of the discrete total variation (DTV) statistic
$d=\sum_i|X_{i+1}-X_i|$ of a multinomial histogram with $n$ ordered cells and $N$
observations, of its version $d_{\mathbf r}$ centered at a null $\mathbf r$ with all
$r_i>0$, and of the cyclic statistic $d^c$; their means and variances are known from
earlier papers. For fixed $n$ and $N\to\infty$, $d/\sqrt N$ under the uniform null
tends, up to the scale $n^{-1/2}$, to the normal-theory sum $D$ of absolute successive
differences, whose first four moments were found by Kamat (1953), with one constant
evaluated numerically. For a fixed null with all $r_i>0$ the raw statistic, after removal of its deterministic drift and division by $\sqrt N$, has a folded (uniform null), a linear Gaussian (no flat edge) or a mixed limit, and one theorem covers all three. We evaluate Kamat's remaining constant
in closed form, so that the cumulants of the uniform limit through order four are
elementary (square roots, $\pi$, $\arcsin\frac13$, $\arcsin\frac23$,
$\arccos\frac14$), and a connected-cluster argument shows that the cumulant of order $m$
of $D$ is affine in $n$ for $n\ge m$. The skewness and the excess kurtosis of the limit law behave like
$1.5218/\sqrt n$ and $2.8677/n$, the factors $0.8370$ and $0.5783$ of the values of the
moment-matched gamma law. For the limit of the cyclic statistic these formulas are exact above small thresholds. Under a general positive null the limit of $d_{\mathbf r}/\sqrt N$ is
a moving average corrected by one common Gaussian factor (a rank-one downdate of its covariance); its third cumulant is in closed form, and for each fixed smooth positive profile $f$ its skewness behaves
like $1.5218\,I_{3/2}/\sqrt n$ with
$I_{3/2}=\int_0^1f^{3/2}\big/\bigl(\int_0^1f\bigr)^{3/2}\ge1$, so among smooth positive
profiles the uniform one minimises the leading skewness; the convergence is not uniform in $f$, and at a given $n$ a comb-like null can be less skewed than the uniform one. For $n=3$ the limit
distribution function is a bivariate normal rectangle probability, and a product of
two error functions when $r_1=r_3$. The uniform limit law has a Gaussian extreme
tail, so an exponential gamma tail must eventually overshoot. Exact integer tables
indicate, without proof for $n\ge3$, an $O(1/N)$ approach of the skewness and the excess kurtosis; for $n=2$ it is proved, with coefficients that depend on the parity of $N$.
A gamma-based Cornish--Fisher quantile, with the limiting skewness and kurtosis and the exact mean and variance, has a median absolute error of $0.26$--$0.32$ of a lattice step at $\alpha=5\%,1\%,0.1\%$ on the tested uniform grid ($0.25$--$0.36$ with the exact finite-$N$ skewness and kurtosis), but used as a critical value with a half-span rounding rule it can be anti-conservative at small $n$ (up to $1.47\alpha$ at
$n=2$). For $n\ge5$ the plain normal Cornish--Fisher quantile with the limiting cumulants is the better critical value at $\alpha\ge10^{-3}$, and the gamma-based one at $\alpha\le10^{-4}$. We give the range where the exact law should be used, and a full-span variant
that was conservative at the three tested levels on the uniform grid for $n\ge5$ but has
no level guarantee.
\end{abstract}



\section{Introduction}
	\label{sec:intro}

Let $X=(X_1,\ldots,X_n)\sim\Mult(N;\mathbf p)$ be the counts of a histogram of $N$
observations over $n$ ordered cells, and let $\mathbf r$ be the null law, so that
$\mathbf p=\mathbf r$ under the null. The linear and cyclic DTV statistics are
\[
	d=\sum_{i=1}^{n-1}|X_{i+1}-X_i|,
	\qquad
	d^c=\sum_{i=1}^{n}|X_{i+1}-X_i|,\qquad X_{n+1}=X_1 ,
\]
and under a general null $\mathbf r$ one also uses the statistic centered at the null,
\[
	d_{\mathbf r}=\sum_{i=1}^{n-1}|X_{i+1}-X_i-N(r_{i+1}-r_i)| ,
\]
so that $d_{\mathbf u}=d$ for the uniform null $\mathbf u$. This is the notation
of~\cite{banic_elezovic_p3,banic_elezovic_p4}, where $d_{\mathbf r}$ is also studied
under a sampling law $\mathbf p\ne\mathbf r$ (there the case $\mathbf p=\mathbf r$ is
called the matched statistic $d_{\mathbf p}$). The present paper is about null laws, so
$\mathbf p=\mathbf r$ throughout.

The statistic $d$ is the statistic of the comb test of uniformity, introduced and
illustrated in~\cite{banic_elezovic_ct}; there its exact null law was computed by
dynamic programming, and a gamma approximation was recommended for the finite
statistic after $29$ candidate families had been fitted to the exact distribution
functions by a discrete Cram\'er--von Mises criterion and compared. The centered statistic $d_{\mathbf r}$ for a
non-uniform null, its exact null law by a weighted dynamic programme, and the sign
locking of the raw statistic $d$ under a non-uniform null were introduced
in~\cite{banic_elezovic_gct}. (For integer weights $r_i=q_i/L$, \cite{banic_elezovic_gct}
works with the integer statistic $L\,d_{\mathbf r}$, writes $g_i$ for the slopes
$\Delta_i=r_{i+1}-r_i$ of Section~\ref{sec:limit}, and $d_{\mathbf u}$ for the
uncentered $d$.) In our earlier work~\cite{banic_elezovic_tvor} the DTV was used as an
outlier measure for histograms. The ideas and illustrations are in these papers; here
we give the mathematical development of the null laws of these statistics.

The first two moments were obtained in the companion parts of this series. The
expectation under the uniform null is
in~\cite{banic_elezovic_p1}, the variance in~\cite{banic_elezovic_p2}, the general
null in~\cite{banic_elezovic_p3,banic_elezovic_p4}, and the cyclic statistic
in~\cite{banic_elezovic_p5}; the sparse regime $N/n\to0$ is treated
in~\cite{elezovic_sparse}. Two statistics with
the same mean and variance may still differ in the right-tail quantiles that fix
critical values. The present paper therefore studies the whole null law: its
discrete structure, its limit, the third and fourth cumulants of the limit, and the
quality of the moment-matched gamma approximation in the tail.

\paragraph{Relation to the literature.}
Under the uniform null the limit of $d/\sqrt N$ is $n^{-1/2}D$, where
$D=\sum_{j=1}^{n-1}|W_{j+1}-W_j|$ is the sum of absolute successive differences of an
i.i.d.\ standard normal sample (Theorem~\ref{thm:limit}(c)); $D/(n-1)$ is the mean
successive difference, or mean moving range, of classical statistics. Its squared
analogue is the statistic of von Neumann, Kent, Bellinson and
Hart~\cite{vonneumann_kent1941}. The mean, the variance and the efficiency of the mean
moving range were obtained by Hoel~\cite{hoel1946}, and also by
Guest~\cite{guest1951} and others (see~\cite{kamat1953}); a recent decomposition of the variance
of the moving-range estimator by the order of the observations is in~\cite{karl2026}.
Kamat~\cite{kamat1953} computed the first
four moments of $D$ from the bivariate and trivariate absolute normal moments of
Nabeya~\cite{nabeya1951,nabeya1952} and of his own paper~\cite{kamat1953a}: the third
moment in closed form, and the fourth up to the absolute moment of four consecutive
differences, which he reduced to elementary terms and one integral, evaluated by
quadrature as $0.328988$; his resulting value for the four-edge term is $2.215484$ (in
our normalisation $4\E|Z_1Z_2Z_3Z_4|$, see Remark~\ref{rem:status}). He tabulated $\beta_1$ and $\beta_2$ for $3\le n\le10$ and
six larger $n$ up to $50$,
gave the exact distribution function for $n=3$ through bivariate normal integrals, and
approximated the law by Pearson Type~I curves fitted to four moments, for $n>8$ by a Type~III (shifted gamma) curve
fitted to three moments; Sathe and Kamat~\cite{sathe_kamat1957} fitted a power of a
$\chi^2$ variable to three moments instead. In~\cite{kamat1953a} Kamat also gave the
first four moments of weighted sums of two or three correlated folded normals and the
distribution function of $|x_1|+c|x_2|$ by bivariate normal integrals, which cover
the general null for $n\le4$ and the distribution function for $n=3$ (see
Section~\ref{ssec:profile}). Nabeya~\cite{nabeya1961} later reduced four-variate
absolute normal moments to elementary terms and one orthant-type integral. Up to the
scale $n^{-1/2}$, the moments of our uniform limit law are therefore Kamat's; our
contribution for it is not its first four moments but an exact elementary evaluation
of the one integral $I$ that he evaluated numerically (as $0.328988$), which turns out to
be $\pi^2/30=0.3289868\ldots$ (Remark~\ref{rem:status}), hence of the four-edge moment
(Lemma~\ref{lem:E1234}; this closed form was obtained earlier by Finch~\cite{finch2015}, in a supplement to Mehta and Normand~\cite{mehta_normand1997}, by another route) and of the fourth
cumulant, the cumulant (connected-cluster)
form of all four (a rewriting of his exact moment formulas~\cite[eq.~(12)]{kamat1953}
that extends to every order), and what follows from it below. The
constants of the limiting cumulants are joint moments of folded normals; the
bivariate ones go back to Nabeya~\cite{nabeya1951} (with Sheppard's arcsine
law~\cite{sheppard1899} for the orthant probabilities) and the trivariate ones to
Nabeya~\cite{nabeya1952} and, independently, Kamat~\cite{kamat1953a}. Among statistics
for ordered categories, the discrete
Kolmogorov--Smirnov statistic~\cite{conover1972,pettitt_stephens1977} and the
discrete Cram\'er--von Mises family~\cite{choulakian1994} have limits that are
functionals of the cumulative multinomial residuals, a Gaussian vector with the
covariance of a Brownian bridge sampled at the cell boundaries, and the separable $L^1$
statistic $\sum_i|X_i-Nr_i|$ of~\cite{gyorfi1991} has, by the multinomial central
limit theorem, as its fixed-$n$ limit a sum of correlated folded normals of the cell
residuals. The DTV limit folds the \emph{increments} of the residuals, with, under the uniform null, the rigid correlation $-\frac12$ between neighbours.

\paragraph{Contributions.}
\begin{enumerate}
\item The support and the parity of the exact law (Proposition~\ref{prop:support}).
\item The identification of the uniform limit with the normal-theory sum of absolute
successive differences, and the convergence of all moments of $d_{\mathbf r}/\sqrt N$
(Theorem~\ref{thm:limit}; the edge covariance and the weak limit for a general null are
in~\cite[Lemma~2.4 and Corollary~3.5]{banic_elezovic_p4}); one limit theorem for the raw statistic covering the
folded, the linear Gaussian and the mixed regime (Theorem~\ref{thm:regimes}).
\item For the uniform limit law: the cumulants $\kappa_2,\kappa_3,\kappa_4$ of
$\sqrt n\,S=D$ as affine functions of $n$, that is,
$\kappa_m(S)=(2/n)^{m/2}(u_mn+v_m)$ for $n\ge m$ with constants $u_m,v_m$ (with
separate formulas for smaller $n$), by a connected-cluster argument that gives the
same structure at every order $m\ge2$ (Theorem~\ref{thm:cumulants} and
the paragraph after Proposition~\ref{prop:cyclic}). The moments themselves
are Kamat's~\cite{kamat1953}, the fourth with one constant evaluated numerically; our additions are the closed form $\pi^2/30$ of that constant (we have not found it stated), and a short proof of the closed form of the four-edge absolute moment (Lemma~\ref{lem:E1234}; the closed form itself is due to Finch~\cite{finch2015}), so that all constants are
elementary (Table~\ref{tab:constants}); the explicit asymptotic skewness and excess
kurtosis constants
$1.5218$ and $2.8677$ (implicit in Kamat's expansions of the central moments in powers of
$1/(n-1)$~\cite[eq.~(12)]{kamat1953}, where only $\beta_1\to0$ and $\beta_2\to3$ are
stated) and the gamma ratios $R_3^\infty=0.8370$, $R_4^\infty=0.5783$
(Corollary~\ref{cor:asymptotics}); the exactness of these formulas for the cyclic
statistic above the thresholds $n\ge3,4,5$ (Proposition~\ref{prop:cyclic}); and the
Gaussian extreme tail of the limit (Proposition~\ref{prop:tail}).
\item The limit $Q_{\mathbf r}$ of $d_{\mathbf r}/\sqrt N$ under a general null with all
$r_i>0$: a representation by a moving average corrected by one common Gaussian factor
(Proposition~\ref{prop:rank-one}); $\kappa_3(Q_{\mathbf r})$ in closed form for every
such profile (Theorem~\ref{thm:kappa3-profile}; $\kappa_2$ is
in~\cite{banic_elezovic_p4}); the large-$n$ profile law
$\sqrt n\,\gamma_1\to1.5218\int f^{3/2}/(\int f)^{3/2}$,
$n\gamma_2\to2.8677\int f^2/(\int f)^2$ for a smooth positive profile $f$
(Theorem~\ref{thm:profile-law}), by which, among smooth positive profiles, the uniform
null minimises the limits of $\sqrt n\gamma_1$, $n\gamma_2$ and of both gamma ratios
(Corollary~\ref{cor:uniform-min}); and, for $n=3$, the exact law through binomial
distribution functions and the limit law as a bivariate normal rectangle probability
(Proposition~\ref{prop:n3}).
\item Numerical evidence from exact integer tables for an $O(1/N)$ approach to the
limiting skewness and excess kurtosis for $n\ge3$, with a proof and closed-form
coefficients for $n=2$ (Section~\ref{sec:finite-N}), and a gamma-based
Cornish--Fisher quantile with a half-span rounding rule, calibrated as a critical value
on these tables, including where it fails (small $n$), with an exact-law range and a
full-span variant that was conservative at the three tested levels on the uniform grid
for $n\ge5$; it is a quantile approximation, not a level-guaranteed test; certified level control needs the exact law (or at least its upper tail at the proposed critical value) or another rigorous device such as a Monte Carlo test (Section~\ref{sec:tails}).
\end{enumerate}
Statements in theorem environments are proved. Statements labelled
\emph{numerical observation} are read from exact integer tables or from Monte Carlo
simulation and are not proved.

\paragraph{Organisation.}
Section~\ref{sec:exact-finite} gives the discrete structure, Section~\ref{sec:limit}
the limit laws, Section~\ref{sec:cumulants} the limiting cumulants and the tail of
the limit under the uniform null, and in Section~\ref{ssec:profile} under a general
null, Section~\ref{sec:finite-N} the finite-$N$ approach,
Section~\ref{sec:gamma} the gamma diagnostic, and Section~\ref{sec:tails} the
evidence from exact tables and the tail approximation. Section~\ref{sec:discussion}
separates what is proved from what is observed, and open problems are collected in
Section~\ref{sec:open}. Exact computation of the null laws for both nulls is the
subject of the companion paper~\cite{banic_elezovic_p08}, and the power of the tests
that of~\cite{banic_elezovic_p10}. The overlap with~\cite{banic_elezovic_p08} is
intentional: the present paper gives the analytic basis, and~\cite{banic_elezovic_p08}
the exact computational procedure.

\section{The exact finite law and its discrete structure}
	\label{sec:exact-finite}

The exact law of $d$ is the probability mass function $k\mapsto\Pr(d=k)$, with
probability generating function
\begin{equation}\label{eq:dtv-pgf-finite}
	G_{N,\mathbf r}(u)
	=
	\sum_{\substack{x_1+\cdots+x_n=N\\ x_i\ge0}}
	\frac{N!}{x_1!\cdots x_n!}
	r_1^{x_1}\cdots r_n^{x_n}
	u^{\sum_{i=1}^{n-1}|x_{i+1}-x_i|}.
\end{equation}
The exact tables used in this paper are integer counts $n^N\Pr(d=k)$ under the uniform
null. They are computed by the dynamic programme introduced in~\cite{banic_elezovic_ct}
for the uniform null and extended to a non-uniform null and to $d_{\mathbf r}$
in~\cite{banic_elezovic_gct}; the algorithm, with a faster variant, is described in
the companion paper~\cite{banic_elezovic_p08}.

\begin{proposition}[Support and parity]\label{prop:support}
Let $N\ge1$ and let all $r_i>0$.
\begin{enumerate}
\item[(a)] For $n\ge4$ the support of $d$ is $\{0,1,\ldots,2N\}\setminus\{1\}$ if
$n\mid N$, and $\{1,2,\ldots,2N\}$ if $n\nmid N$.
\item[(b)] For $n=3$ the support of $d$ in $[0,N]$ is as in~(a), and above $N$ the
statistic takes exactly the values $N<k\le2N$ with $k\equiv2N\pmod 3$.
\item[(c)] For $n=2$, $d=|2X_1-N|$, so $d\equiv N\pmod2$.
\item[(d)] For $n\ge3$ the support of $d^c$ is $\{0,2,\ldots,2N\}\setminus\{2\}$ if
$n\mid N$, and $\{2,4,\ldots,2N\}$ if $n\nmid N$.
\item[(e)] $d\equiv X_1+X_n\pmod2$. Under the uniform null,
$\Pr(d\text{ odd})=\frac12\bigl(1-(1-4/n)^N\bigr)$.
\end{enumerate}
\end{proposition}

\begin{proof}
A nondecreasing path $u=x_1\le\cdots\le x_n=u+k$ has $d=k$, and by raising one
coordinate at a time its total takes every integer value in
$[nu+k,\,nu+(n-1)k]$. Hence $k\le N$ is attained by a monotone path if and only if
some multiple $nu\ge0$ lies in $[N-(n-1)k,\,N-k]$. For $n\ge3$ and $k\ge2$ this
interval has length $(n-2)k\ge n-1$, so it contains a multiple of $n$ (or contains
$0$ if its left end is negative). For $k=1$ the interval $[N-n+1,N-1]$ contains a
multiple of $n$ if and only if $n\nmid N$, and $d=1$ forces a monotone path with a
single unit step; $d=0$ forces a constant path, which needs $n\mid N$. For $n\ge4$ the
values $N\le k\le2N$ come from $x=(0,\ldots,0,j,0,N-j)$, with $j$ in position $n-2$,
which has $d=N+j$, $0\le j\le N$. Since $|x_{i+1}-x_i|\le x_i+x_{i+1}$, always
$d\le2N$ and $d^c\le2N$. This proves~(a). For $n=3$ write $x=(a,b,c)$. A
monotone path has $d=|c-a|\le N$, a valley at $b$ has $d=N-3b\le N$, and a peak
at $b$ has $d=3b-N$, which exceeds $N$ exactly when $b>2N/3$; conversely, for
$N<k\le2N$ with $k\equiv2N\pmod3$ the point $(N-b,b,0)$, $b=(N+k)/3$, has $d=k$. This
gives~(b). Part~(c) is immediate. For~(d) note that $d^c$ is twice the total ascent
around the cycle; a monotone path closed into a cycle has $d^c=2k$ with the same
totals as above, and the argument of~(a) applies to $k\in[0,N]$. Here $d^c=0$ forces
a constant vector, and $d^c=2$ forces, up to rotation, one block of entries $u+1$ and
the rest equal to $u$, whose total is not a multiple of $n$; conversely such a vector
exists for every $N$ with $n\nmid N$. Finally $|m|\equiv m\pmod 2$
gives $d\equiv\sum_i(X_{i+1}-X_i)=X_n-X_1$, and $X_1+X_n\sim\mathrm{Bin}(N,2/n)$
gives~(e).
\end{proof}

The cyclic law lives on the even lattice, while the linear law has no parity
constraint for $n\ge3$: for $n=4$ the parity is exactly balanced, and for $n\ge5$ it is
balanced up to $(1-4/n)^N/2$. We checked
Proposition~\ref{prop:support} by exhaustive enumeration for $n\le6$ and small $N$.
For the centered statistic $d_{\mathbf r}$ with $r_i=q_i/L$ the analogous facts are
different: $d_{\mathbf r}=0$ is possible if and only if $L\mid N$ (for
$\gcd_iq_i=1$), and, for $n\ge3$, the support of $d_{\mathbf r}$ lies in one coset of
$(s/L)\mathbb Z$, where $s=g$ if $L/g$ is odd and $s=2g$ if $L/g$ is even, with
$g=\gcd\bigl(L,N\gcd_i(q_{i+1}-q_i)\bigr)$ (and $\gcd_i(q_{i+1}-q_i)=0$ if all $q_i$
are equal); the step is exactly $s/L$ for
$N\ge N_0=(4n-2)L/\min_iq_i$~\cite[Theorem~3.1]{banic_elezovic_p08}, and it may fail
below $N_0$: for $\mathbf q=(1,8,1)$ and $N=3$ the support of $L\,d_{\mathbf r}$ is
$\{12,18,42,72\}$, of step $6$, while $s=2$. (For $n=2$, $L\,d_{\mathbf r}=2|LX_1-Nq_1|$ has its
own step, see~\cite{banic_elezovic_p08}.)

\begin{observation}[Shape on the grid]\label{obs:shape}
On the uniform grid of Section~\ref{ssec:grid} ($13\times18=234$ cells, $2\le n\le50$,
$10\le N\le500$) the holes of the exact law are exactly those of
Proposition~\ref{prop:support}: one hole at $d=1$ or at $d=0$ for $n\ge4$, the band
structure above $N$ for $n=3$, and the deterministic parity for $n=2$. The law is
\emph{not} unimodal in general. Counting local maxima on the support whose height is
at least $10^{-3}$ of the mode (a local maximum is a support point whose mass strictly
exceeds that at each adjacent support point; at an endpoint only the one neighbour is
used):
\begin{enumerate}
\item[(a)] for $n=3$ every cell has many local maxima: the probability mass function
is a permanent sawtooth of period $3$ (at $N=300$ there are $30$ local maxima above
$10^{-3}$ of the mode, at $15$ and $21$ with heights $0.99$ and $0.96$ of the mode
at $18$);
\item[(b)] for $n=4$ every cell, and for $n=6$ twelve of the $18$ cells, have
secondary maxima, near the mode often within a few per cent of its height (at
$N=300$, $n=4$: heights $0.98$
and $0.94$ at $19$ and $27$, mode $23$; at $N=200$, $n=6$: height $0.99$ at $27$, mode
$29$);
\item[(c)] for $n=2,8,9,12$ every cell has a single local maximum above this
threshold, and for $n=5,7,10,20,30,50$ this holds once $N\ge40,20,20,20,30,60$,
respectively. For $n\ge10$ and smaller $N$ the secondary maxima alternate with step
$2$; this is consistent with the parity imbalance of
Proposition~\ref{prop:support}(e), which is visible while $(1-4/n)^N$ is not small (a
heuristic, not a proof). For $n=5$ and $N\le30$ the secondary maximum next to the mode
lies $3$ below it (at $N=10$ there is also a small tail maximum at $d=17$, of
relative height $0.008$), and for $n=7$, $N=10$ there is a tail bump at $d=2N$; we have no mechanism for these.
\end{enumerate}
Below $10^{-3}$ of the mode there are further local maxima in the far tail (apart from
the continuing $n=3$ sawtooth, for $n=5,7,9,30,50$; e.g.\ at $d=2N=40$ for $N=20$,
$n=9$), so the law is unimodal in the strict sense only in some of these cells (for
$n=2,8,12$ in all cells of the grid).
\end{observation}

The period-$3$ sawtooth is proved, as a local limit law with residue-dependent densities, in~\cite[Theorem~3.2]{banic_elezovic_p08}. The mechanism: for $n=3$,
$d=\max(|X_3-X_1|,|3X_2-N|)$ exactly (Proposition~\ref{prop:n3}), and the second term
only takes values $\equiv\pm N\pmod 3$, so the three residue classes of $d$ modulo $3$
can receive different masses. We have no mechanism for the secondary maxima at $n=4$ and
$n=6$.

\section{Limit laws}
	\label{sec:limit}

Write $D_i=X_{i+1}-X_i$, $\Delta_i=r_{i+1}-r_i$ and $\tilde D_i=D_i-N\Delta_i$, so
that $d_{\mathbf r}=\sum_i|\tilde D_i|$.

\begin{theorem}[Edge-vector limit and the folded limit law]\label{thm:limit}
Fix $n\ge2$ and $\mathbf r$ with all $r_i>0$, and let $N\to\infty$.
\begin{enumerate}
\item[(a)] \textup{(\cite[Lemma~2.4 and Corollary~3.5]{banic_elezovic_p4})}
$N^{-1/2}(\tilde D_1,\ldots,\tilde D_{n-1})$ converges in distribution to a
centered Gaussian vector $\mathbf G^{(\mathbf r)}$ with
\begin{gather*}
	\Var G_i^{(\mathbf r)}=r_i+r_{i+1}-\Delta_i^2,\qquad
	\Cov(G_i^{(\mathbf r)},G_{i+1}^{(\mathbf r)})=-r_{i+1}-\Delta_i\Delta_{i+1},\\
	\Cov(G_i^{(\mathbf r)},G_j^{(\mathbf r)})=-\Delta_i\Delta_j\quad (|i-j|\ge2).
\end{gather*}
\item[(b)] $d_{\mathbf r}/\sqrt N\to Q_{\mathbf r}:=\sum_{i=1}^{n-1}|G_i^{(\mathbf r)}|$
in distribution and with convergence of all moments.
\item[(c)] Under the uniform null, $\Var G_i=2/n$, $\Cov(G_i,G_{i+1})=-1/n$ and
$\Cov(G_i,G_j)=0$ for $|i-j|\ge2$. Equivalently,
\[
	(G_1,\ldots,G_{n-1})\overset{d}{=}n^{-1/2}(W_2-W_1,\ldots,W_n-W_{n-1}),
	\qquad W_1,\ldots,W_n\ \text{i.i.d.\ }\mathcal N(0,1),
\]
so that
\begin{equation}\label{eq:S-msd}
	\frac{d}{\sqrt N}\ \xrightarrow{\ d\ }\ S:=\sum_{j=1}^{n-1}|G_j|
	\overset{d}{=}\frac1{\sqrt n}\sum_{j=1}^{n-1}|W_{j+1}-W_j| .
\end{equation}
For the cyclic statistic, $d^c/\sqrt N\to S_c:=\sum_{j=1}^n|G_j|$, with the same
representation and $W_{n+1}=W_1$.
\end{enumerate}
\end{theorem}

\begin{proof}
(a) The vector $(\tilde D_i)$ is a fixed linear image $a_i^{\!\top}(X-N\mathbf r)$,
$a_i=e_{i+1}-e_i$, of the centered count vector, whose normalized covariance is
$\operatorname{diag}(\mathbf r)-\mathbf r\mathbf r^{\!\top}$; since $X$ is a sum of
$N$ i.i.d.\ one-hot vectors, the multivariate central limit
theorem~\cite[Section~29]{billingsley} gives a Gaussian limit with covariance
$a_i^{\!\top}(\operatorname{diag}\mathbf r-\mathbf r\mathbf r^{\!\top})a_j$, which
evaluates to the three displayed cases.
(b) Convergence in distribution follows from (a) by the continuous mapping theorem,
as in~\cite[Corollary~3.5]{banic_elezovic_p4}. Each $\tilde D_i$ is a sum of $N$
i.i.d.\ centered increments bounded by $2$, so $\E|\tilde D_i/\sqrt N|^k$ is bounded in
$N$ for every $k$. Since $d_{\mathbf r}/\sqrt N=\sum_i|\tilde D_i|/\sqrt N$, every
power of $d_{\mathbf r}/\sqrt N$ is uniformly integrable, and all moments converge.
(c) The covariances are the case $r_i=1/n$ of (a). The vector
$Y=n^{-1/2}(W-\bar W\mathbf 1)$ has covariance $n^{-1}(I-n^{-1}\mathbf 1\mathbf
1^{\!\top})$, the normalized multinomial covariance under the uniform null, and the
differences $Y_{j+1}-Y_j=n^{-1/2}(W_{j+1}-W_j)$ do not contain $\bar W$.
\end{proof}

Under the uniform null $Q_{\mathbf r}=S$. For a non-uniform null the edge covariance
of part~(a) need not be tridiagonal: non-adjacent edges are correlated through
$-\Delta_i\Delta_j$, which vanishes for all non-adjacent pairs only for special
profiles (for instance $\mathbf r=(1,2,2,2)/7$, where only $\Delta_1\ne0$), and the
edge variances are in general not constant along the edges. The stationary cluster
formulas of Section~\ref{ssec:uniform-cumulants} therefore do not extend verbatim; a general null is reduced to a one-dependent structure in
Section~\ref{ssec:profile} by a representation with one common factor
(Proposition~\ref{prop:rank-one}). The variance of $Q_{\mathbf r}$ is the leading variance
of $d_{\mathbf r}$~\cite[Corollary~3.5]{banic_elezovic_p4} (expanded in~\cite[Theorem~3.3]{banic_elezovic_p4} for $n\ge5$ and in~\cite[Corollary~6.4]{banic_elezovic_p4} for $n\ge3$); under the uniform null the mean and
variance of $S$ are the leading terms of the expansions of $\E d$
in~\cite[Lemma~5.1 and Theorem~5.5]{banic_elezovic_p1} (stated there for $d^c$; under the uniform null $\E d=\frac{n-1}n\E d^c$~\cite[Section~3.1]{banic_elezovic_p1}) and of $\Var d$
in~\cite[Theorem~3.4 and Corollary~3.5]{banic_elezovic_p2}.

\begin{remark}[The density]\label{rem:density}
For $n=3$ the density is explicit in error functions. With $S=\sqrt{2/3}\,S_3$, $S_3=|U|+|V|$, $U,V$
standard normal and $\Corr(U,V)=-\frac12$, integration of the bivariate normal density along the sides of
the square $|u|+|v|=s$ gives
\[
	f_{S_3}(s)=\sqrt{\tfrac2\pi}\,e^{-s^2/2}\operatorname{erf}\!\bigl(s/\sqrt6\bigr)
	+\sqrt{\tfrac2{3\pi}}\,e^{-s^2/6}\operatorname{erf}\!\bigl(s/\sqrt2\bigr),
	\qquad s>0 .
\]
(We checked the mass, the mean $2\sqrt{2/\pi}$, the variance $2A+2B$ ($A,B$ as in
Table~\ref{tab:constants}) and the standardized cumulants $1.0176156$ and $1.2962273$
of Table~\ref{tab:gamma-n} by quadrature.) For $n=4$ the same method leads to bivariate
normal probabilities of triangles, which reduce to combinations of Owen's $T$
function~\cite{owen1956}. We do not pursue the density for general $n$ and work with
cumulants instead. For $n=3$ and any null the distribution function is a bivariate
normal rectangle probability (Proposition~\ref{prop:n3}; cf.~\cite[\S4.2]{kamat1953a}),
a product of two error functions when $r_1=r_3$; under the uniform null it is
$\Pr(S\le x)=\operatorname{erf}(\sqrt3\,x/2)\operatorname{erf}(x/2)$ for $x\ge0$, whose
derivative is the density above after the change of scale $S=\sqrt{2/3}\,S_3$.
Kamat~\cite[eq.~(13)]{kamat1953} gave this distribution function as a combination of
bivariate normal integrals; the product form is equivalent to it, since
$S_3=\max(|U+V|,|U-V|)$ with $U+V$ and $U-V$ independent.
\end{remark}

\begin{theorem}[Regimes of the raw statistic]\label{thm:regimes}
Fix $n\ge2$ and $\mathbf r$ with all $r_i>0$; let $I_0=\{i:\Delta_i=0\}$,
$I_1=\{i:\Delta_i\ne0\}$ and $\mathbf G=\mathbf G^{(\mathbf r)}$. Then
\begin{equation}\label{eq:mixed-limit}
	\frac{d-N\sum_{i\in I_1}|\Delta_i|}{\sqrt N}
	\;\xrightarrow{d}\;
	\sum_{i\in I_0}|G_i|
	+
	\sum_{i\in I_1}\sgn(\Delta_i)\,G_i .
\end{equation}
\end{theorem}

\begin{proof}
For $i\in I_0$, $|D_i|=|\tilde D_i|$. For $i\in I_1$, Hoeffding's
inequality~\cite{hoeffding1963} gives
$\Pr\{\sgn D_i\ne\sgn\Delta_i\}\le e^{-N\Delta_i^2/2}$, as in the proof
of~\cite[Theorem~6.5]{banic_elezovic_p3}. On the complementary event
$|D_i|=N|\Delta_i|+\sgn(\Delta_i)\tilde D_i$ \emph{exactly}. Hence, with probability
tending to one,
\[
	d-N\sum_{i\in I_1}|\Delta_i|=\sum_{i\in I_0}|\tilde D_i|+\sum_{i\in I_1}\sgn(\Delta_i)\tilde D_i ,
\]
and the right-hand side divided by $\sqrt N$ converges by
Theorem~\ref{thm:limit}(a) and the continuous mapping theorem.
\end{proof}

The convergence in~\eqref{eq:mixed-limit} holds with all moments: since
$\bigl||D_i|-N|\Delta_i|\bigr|\le|\tilde D_i|$, the left-hand side is at most
$\sum_i|\tilde D_i|/\sqrt N$ in absolute value, whose moments are bounded in $N$ (proof of
Theorem~\ref{thm:limit}(b)), so every power of it is uniformly integrable. The limit is
taken at fixed $\mathbf r$ and is not uniform in $\mathbf r$: the bound
$e^{-N\Delta_i^2/2}$ is small only when $N\Delta_i^2$ is large. Near a tie, with
$\sqrt N\Delta_i\to\lambda_i$ finite along a sequence of nulls, the same argument gives
$|D_i|/\sqrt N\to|G_i+\lambda_i|$, a folded normal with nonzero mean, which neither term
of~\eqref{eq:mixed-limit} describes; for nulls with nearly equal neighbouring $r_i$ the
fixed-$\mathbf r$ limit is therefore a poor guide at practical $N$.

If $I_1=\varnothing$, \eqref{eq:mixed-limit} is the folded limit of
Theorem~\ref{thm:limit}(b). If $I_0=\varnothing$, the limit is a centered normal
variable, and $d/\sqrt N\to\infty$ in probability. This linear regime is the sign locking
introduced in~\cite{banic_elezovic_gct}, where it is shown that the raw $d$ then
reduces to a fixed signed sum of the counts and loses phase-blindness; the near-normal
shape of the raw statistic under a non-uniform null is illustrated there by simulation.
Theorem~\ref{thm:regimes} gives the limit law in all three regimes. The variance in
the linear regime is~\cite[Theorem~5.5]{banic_elezovic_p4}, and the mixed regimes for the mean and the
variance under local alternatives are~\cite[Corollary~6.9]{banic_elezovic_p3}
and~\cite[Proposition~5.7]{banic_elezovic_p4}.

\begin{remark}[Finite-$N$ shape in the linear regime]\label{rem:linear-cumulants}
If $I_0=\varnothing$, then off an event of probability at most $\sum_ie^{-N\Delta_i^2/2}$ (proof of Theorem~\ref{thm:regimes})
$d=\sum_k c_kX_k$ with $c_k=\sgn\Delta_{k-1}-\sgn\Delta_k$ (terms with an index outside
$1,\ldots,n-1$ omitted). The cumulants of a linear multinomial statistic are
$\kappa_m=N\kappa_m(c_J)$ with $\Pr(J=k)=r_k$, so
\[
	\gamma_1=\frac{\kappa_3(c_J)}{\kappa_2(c_J)^{3/2}\sqrt N},\qquad
	\gamma_2=\frac{\kappa_4(c_J)}{\kappa_2(c_J)^2N},
\]
up to exponentially small terms. The limit is normal ($\gamma_1,\gamma_2\to0$), and at
finite $N$ either sign of $\gamma_1$ and $\gamma_2$ can occur. For the profile
$\mathbf r=(1,2,1)/4$ the sign-locked statistic is $3X_2-N$ with
$X_2\sim\mathrm{Bin}(N,\frac12)$, so $\gamma_1=0$ and $\gamma_2=-2/N$ for it, and for $d$
up to exponentially small terms (at $N=3$, $d\ne3X_2-N$ with probability $5/16$).
\end{remark}

\begin{remark}[The other axis: $n\to\infty$]\label{rem:sparse-axis}
The regimes above are regimes of the edge trends at fixed $n$. Under the uniform null,
if $n\to\infty$ with $N/n\to\rho\in(0,\infty)$, a single edge does not become Gaussian: two neighbouring
counts tend to independent Poisson variables with mean $\rho$, and
$X_{i+1}-X_i\to\mathrm{Skellam}(\rho,\rho)$~\cite{irwin1937,skellam1946}. In the
Poissonized model the statistic is then a sum of $n$ one-dependent folded Skellam
terms, and $\E d^c/n\to\varphi(\rho)=\E|\mathrm{Skellam}(\rho,\rho)|
=2\rho e^{-2\rho}\{\mathrm I_0(2\rho)+\mathrm I_1(2\rho)\}$~\cite{ramasubban1958,katti1960},
where $\mathrm I_\nu$ is the modified Bessel function of the first kind. The
normalized curves $\varphi(\rho)/\rho$ and $\varphi(\rho)/\sqrt\rho$ tend to the sparse
constant $2$ as $\rho\to0$ and to the dense constant $2/\sqrt\pi$ as $\rho\to\infty$,
respectively. For separable statistics the fluctuations are Gaussian on the scale
$\sqrt n$, with a variance reduced by the conditioning on the total count, by the
theory of~\cite{holst1972,morris1975}; in the Poissonized model the cyclic DTV is a sum of one-dependent terms,
and its central limit theorem along the ray is not treated here. In terms of the effective count
$\Lambda=Ns^2$, where $s^2=2/n$ is the variance of one
edge increment, the fixed-$n$ setting of the present paper lies in $\Lambda\to\infty$,
the ray regime is $\Lambda\to2\rho\in(0,\infty)$, and $\Lambda\to0$ is the sparse
regime of~\cite{elezovic_sparse}.
\end{remark}

\section{Limiting cumulants}
	\label{sec:cumulants}

\subsection{The uniform null}
	\label{ssec:uniform-cumulants}

Let $\sigma^2=2/n$, $Z_j=G_j/\sigma$ (so $\Corr(Z_j,Z_{j+1})=-\frac12$), $c=\sqrt{2/\pi}$
and $H_j=|Z_j|-c$. We write $\kappa(\cdot,\ldots,\cdot)$ for joint cumulants, and
$\Phi$ and $\varphi$ for the standard normal distribution function and density.

\begin{lemma}[Connected clusters]\label{lem:clusters}
A joint cumulant $\kappa(H_{i_1},\ldots,H_{i_m})$ vanishes unless the set
$\{i_1,\ldots,i_m\}$ is a set of consecutive integers.
\end{lemma}

\begin{proof}
By~\eqref{eq:S-msd}, $H_j$ is a function of $(W_j,W_{j+1})$. If the index set is not
an interval, it splits into two nonempty groups that use disjoint sets of the $W$'s;
the two groups are independent, and a joint cumulant of two independent groups
vanishes~\cite{leonov_shiryaev1959}.
\end{proof}

Counting the ordered index tuples whose set is an interval of length $L$ (there are
$n-L$ such intervals among the $n-1$ edges) gives the following theorem. The
constants are the joint cumulants listed in Table~\ref{tab:constants}; translation
invariance makes them independent of the position, and reflection gives
$\kappa(H_1,H_1,H_2,H_3)=\kappa(H_1,H_2,H_3,H_3)=b_{\mathrm{end}}$.

\begin{theorem}[Cumulants of the limit law]\label{thm:cumulants}
Under the uniform null, $\E S=(n-1)c\,\sigma=2(n-1)/\sqrt{\pi n}$ and
\begin{align}
	\kappa_2(S)&=\sigma^2\{(n-1)A+2(n-2)B\}=\sigma^2\{(n-1)(A+2B)-2B\}, && n\ge2,
	\label{eq:kappa2-limit}\\
	\kappa_3(S)&=\sigma^3\{(n-1)a_3+6(n-2)c_{21}+6(n-3)c_{111}\}
	=\sigma^3\{(n-1)C-e_3\}, && n\ge3,
	\label{eq:kappa3-limit}\\
	\kappa_4(S)&=\sigma^4\{(n-1)b_1+(n-2)(8b_{31}+6b_{22})\notag\\
	&\qquad+(n-3)(12b_{121}+24b_{\mathrm{end}})+24(n-4)b_{1111}\}\notag\\
	&=\sigma^4\{(n-1)C_4-e_4\}, && n\ge4,
	\label{eq:kappa4-limit}
\end{align}
where
\[
	C=a_3+6c_{21}+6c_{111},\quad e_3=6c_{21}+12c_{111},\quad
	C_4=b_1+8b_{31}+6b_{22}+12b_{121}+24b_{\mathrm{end}}+24b_{1111},
\]
$e_4=(8b_{31}+6b_{22})+2(12b_{121}+24b_{\mathrm{end}})+72b_{1111}$, and the constants
are given in Table~\ref{tab:constants}. For smaller $n$ the groups with a
negative multiplicity are omitted: $\kappa_3(S)=\sigma^3a_3$ for $n=2$, and
$\kappa_4(S)=\sigma^4b_1$ for $n=2$, $\kappa_4(S)=\sigma^4(2b_1+8b_{31}+6b_{22})$ for
$n=3$.
\end{theorem}

The proof is given after Lemma~\ref{lem:E1234} below: the constants of four
consecutive edges need the closed form of one four-edge absolute moment, which is
proved first.

\begin{table}[ht]
\centering\small
\caption{The constants of Theorem~\ref{thm:cumulants}. All joint cumulants are of the
centered folded variables $H_j=|Z_j|-\sqrt{2/\pi}$ on consecutive edges, with
$\Corr(Z_j,Z_{j+1})=-\frac12$.}
\label{tab:constants}
\footnotesize
\setlength{\tabcolsep}{4pt}
\renewcommand{\arraystretch}{1.35}
\begin{tabular}{llll}
\toprule
& cumulant & closed form & value\\
\midrule
$A$ & $\kappa(H_1,H_1)$ & $1-\frac2\pi$ & $0.36338022763241865692$\\
$B$ & $\kappa(H_1,H_2)$ & $\frac{\sqrt3-2}{\pi}+\frac16$ & $0.08137578972087737310$\\
$a_3$ & $\kappa_3(H_1)$ & $\frac{\sqrt2(4-\pi)}{\pi^{3/2}}$ & $0.21801361414499016069$\\
$c_{21}$ & $\kappa(H_1,H_1,H_2)$ & $\frac{\sqrt2(48-24\sqrt3-\pi)}{12\pi^{3/2}}$ & $0.06961416771785920292$\\
$c_{111}$ & $\kappa(H_1,H_2,H_3)$ & $\pi^{-3/2}\bigl[2+2\sqrt2-2\sqrt6+\frac{\sqrt2}6\pi-\frac{3\sqrt2}2\arcsin\frac13\bigr]$ & $-0.00915442134040488128$\\
$b_1$ & $\kappa_4(H_1)$ & $\frac8\pi-\frac{24}{\pi^2}$ & $0.11477068205421885765$\\
$b_{31}$ & $\kappa(H_1,H_1,H_1,H_2)$ & $\pi^{-2}\bigl[12\sqrt3-24+(\frac52-\frac{3\sqrt3}4)\pi\bigr]$ & $0.05649089189557401254$\\
$b_{22}$ & $\kappa(H_1,H_1,H_2,H_2)$ & $\pi^{-2}\bigl[16\sqrt3-30+(\frac23-\frac{2\sqrt3}3)\pi+\frac{4\pi^2}9\bigr]$ & $0.05735793624037422979$\\
$b_{121}$ & $\kappa(H_1,H_2,H_2,H_3)$ & $\pi^{-2}\bigl[16\sqrt3-22-4\sqrt2+(\frac23-\frac{2\sqrt3}3)\pi-\frac{\pi^2}{18}+6\arcsin\frac13\bigr]$ & $0.00136430459579375027$\\
$b_{\mathrm{end}}$ & $\kappa(H_1,H_1,H_2,H_3)$ & $\pi^{-2}\bigl[8\sqrt3-8-4\sqrt2+(\frac{\sqrt3}4-\frac76)\pi+6\arcsin\frac13\bigr]$ & $-0.00671437633526067005$\\
$b_{1111}$ & $\kappa(H_1,H_2,H_3,H_4)$ & see \eqref{eq:b1111} & $0.00115605405050722623$\\
\midrule
$C$ & & $\pi^{-3/2}\bigl[12+40\sqrt2-24\sqrt6-\frac{\sqrt2}2\pi-9\sqrt2\arcsin\frac13\bigr]$ & $0.58077209240971609051$\\
$C_4$ & & see \eqref{eq:C4} & $0.79381735497649868826$\\
\bottomrule
\end{tabular}
\end{table}

The constants involving four edges are
\begin{align}
	\pi^2b_{1111}&=6\sqrt3-7-4\sqrt2+\sqrt5-\Bigl(\frac32+\frac{\sqrt3}3\Bigr)\pi+\frac{\pi^2}{180}
	\notag\\
	&\qquad+6\arcsin\tfrac13+3\arcsin\tfrac23+\sqrt3\arccos\tfrac14,
	\label{eq:b1111}\\
	\pi^2C_4&=720\sqrt3-1020-240\sqrt2+24\sqrt5-(24+20\sqrt3)\pi+\frac{32\pi^2}{15}
	\notag\\
	&\qquad+360\arcsin\tfrac13+72\arcsin\tfrac23+24\sqrt3\arccos\tfrac14 .
	\label{eq:C4}
\end{align}

\begin{lemma}[The four-edge absolute moment]\label{lem:E1234}
\begin{equation}\label{eq:EZ1234}
	\E|Z_1Z_2Z_3Z_4|=\frac{\sqrt5+\frac{\pi^2}{30}-\frac\pi2+3\arcsin\frac23+\sqrt3\arccos\frac14}{\pi^2}
	=0.55387077505637946715\ldots .
\end{equation}
\end{lemma}

\begin{proof}
In the representation~\eqref{eq:S-msd}, $Z_j=(W_{j+1}-W_j)/\sqrt2$. Given $W_3=w$, the
pairs $(W_1,W_2)$ and $(W_4,W_5)$ are independent and identically distributed, so
$\E|Z_1Z_2Z_3Z_4|=\frac14\E[h(W_3)^2]$ with
$h(w)=\E\bigl(|W_2-W_1|\,|W_3-W_2|\bigm|W_3=w\bigr)$. With
$\E|x-W_1|=2\varphi(x)+x(2\Phi(x)-1)$, one integration by parts gives
\[
	h(w)=a(w)+b(w)+\psi(w),\qquad
	a(w)=\frac{2w}{\sqrt\pi}\operatorname{erf}(w),\quad
	b(w)=\frac2\pi e^{-w^2},\quad
	\psi(w)=2\Phi(w)\bigl(1-\Phi(w)\bigr).
\]
Let $W\sim\mathcal N(0,1)$. We use three tools: $\E e^{-sW^2}=(1+2s)^{-1/2}$; Stein's
identity $\E[Wg(W)]=\E g'(W)$, which removes the polynomial factors; and the orthant
formula $\E(2\Phi(\alpha W)-1)(2\Phi(\beta W)-1)=\frac2\pi\arcsin\bigl(\alpha\beta/
\sqrt{(1+\alpha^2)(1+\beta^2)}\bigr)$ (Sheppard~\cite{sheppard1899}), with
$\operatorname{erf}(w)=2\Phi(\sqrt2w)-1$. Moreover $2\Phi(W)-1$ is uniform on $(-1,1)$ and
$\psi=\frac12(1-(2\Phi-1)^2)$, so $\E\psi(W)^2=\frac14(1-\frac23+\frac15)=\frac2{15}$. The six
terms of $\E h(W)^2$ are
\begin{align*}
	\E a^2&=\frac8{\pi^2}\arcsin\frac23+\frac{32}{3\sqrt5\,\pi^2}, &
	\E b^2&=\frac4{\sqrt5\,\pi^2},\qquad \E\psi^2=\frac2{15},\\
	2\E ab&=\frac{16}{3\sqrt5\,\pi^2}, &
	2\E b\psi&=\frac2{\sqrt3\,\pi}-\frac4{\sqrt3\,\pi^2}\arcsin\frac14,\\
	2\E a\psi&=\frac4{\sqrt3\,\pi}-\frac{8}{\sqrt3\,\pi^2}\arcsin\frac14
	-\frac8{\pi^2}\arcsin\frac1{\sqrt6}.
\end{align*}
Summing and dividing by $4$,
\[
	\E|Z_1Z_2Z_3Z_4|=\frac1{30}+\frac{\sqrt3}{2\pi}
	+\frac{\sqrt5+2\arcsin\frac23-2\arcsin\frac1{\sqrt6}-\sqrt3\arcsin\frac14}{\pi^2},
\]
which is~\eqref{eq:EZ1234} by $2\arcsin(1/\sqrt6)=\arccos\frac23=\frac\pi2-\arcsin\frac23$
and $\sqrt3\arcsin\frac14=\frac{\sqrt3\pi}2-\sqrt3\arccos\frac14$.
\end{proof}

\begin{proof}[Proof of Theorem~\ref{thm:cumulants}]
By Lemma~\ref{lem:clusters}, $\kappa_m(S)=\sigma^m\sum\kappa(H_{i_1},\ldots,H_{i_m})$
over ordered tuples whose index set is an interval. For $m=3$ there are $n-1$ tuples
$(i,i,i)$; for each of the $n-2$ adjacent pairs, two choices of the repeated index and
three positions give $6$ ordered tuples; each of the $n-3$ consecutive triples gives
$3!=6$. For $m=4$ an adjacent pair gives $8$ tuples of type $\{i,i,i,i+1\}$ (two
choices of the tripled index, four positions) and $\binom42=6$ of type
$\{i,i,i+1,i+1\}$; a consecutive triple gives $12$ tuples with the middle index
doubled and $24$ with an end index doubled; four consecutive edges give $4!=24$. The
affine forms follow from $(n-L)=(n-1)-(L-1)$.

The constants reduce to absolute moments of at most four consecutive $Z_j$; the
ones of at most three are also in~\cite[eq.~(4)]{kamat1953}. For
two variables with correlation $\rho$, $\E|Z_1Z_2|=\frac2\pi(\sqrt{1-\rho^2}+\rho\arcsin\rho)$
(Nabeya's formula~\cite{nabeya1951}, cf.~\cite[Lemma~3.1]{banic_elezovic_p2}),
$\E Z_1^2|Z_2|=c(1+\rho^2)$ and
$\E|Z_1|^3|Z_2|=\frac2\pi\{(2+\rho^2)\sqrt{1-\rho^2}+3\rho\arcsin\rho\}$; at
$\rho=-\frac12$ these are $\frac{\sqrt3}\pi+\frac16$, $\frac54c$ and
$\frac12+\frac{9\sqrt3}{4\pi}$. Together with the half-normal
moments~\cite{leone1961} they give $A$, $B$, $a_3$, $b_1$, $c_{21}$, $b_{31}$ and
$b_{22}$. For three consecutive edges ($\rho_{12}=\rho_{23}=-\frac12$, $\rho_{13}=0$)
Nabeya's trivariate formulas~\cite{nabeya1952} give
\begin{equation}\label{eq:EZ123}
	\E|Z_1Z_2Z_3|=\Bigl(\frac2\pi\Bigr)^{3/2}\Bigl[\frac1{\sqrt2}+\frac\pi4-\frac34\arcsin\frac13\Bigr]
	=\frac{4+\sqrt2\,\pi-3\sqrt2\arcsin\frac13}{2\pi^{3/2}},
\end{equation}
where the partial correlations $-1/\sqrt3$ and $-\frac13$ enter through
$2\arcsin(1/\sqrt3)=\frac\pi2-\arcsin\frac13$, and
$\E Z_1^2|Z_2Z_3|=\frac16+\frac{5\sqrt3}{4\pi}$,
$\E|Z_1|Z_2^2|Z_3|=\frac3\pi$. Since
$c_{111}=\E|Z_1Z_2Z_3|-c(2\E|Z_1Z_2|+\E|Z_1Z_3|)+2c^3$ with $\E|Z_1Z_3|=c^2$, this gives
$c_{111}$; $b_{121}$ and $b_{\mathrm{end}}$ follow in the same way from
$b_{121}=\E H_1H_2^2H_3-2B^2$ and $b_{\mathrm{end}}=\E H_1^2H_2H_3-AB$. Finally
$b_{1111}=\E H_1H_2H_3H_4-B^2$, in which $\E|Z_1Z_2Z_4|=c\,\E|Z_1Z_2|$ and
$\E|Z_1Z_3Z_4|=c\,\E|Z_3Z_4|$ by independence, and the remaining moment needed here is
$\E|Z_1Z_2Z_3Z_4|$, given by Lemma~\ref{lem:E1234} above.
\end{proof}

Up to the factor $n^{m/2}$, the cumulants of Theorem~\ref{thm:cumulants} are those of
Kamat's sum $D$: his moment
formulas~\cite[eqs.~(4), (11), (12)]{kamat1953} give~\eqref{eq:kappa2-limit}
and~\eqref{eq:kappa3-limit}, and~\eqref{eq:kappa4-limit} once his numerical value of
the four-edge moment~\cite[eq.~(8)]{kamat1953} is replaced by Lemma~\ref{lem:E1234}
(this holds for his closed forms; some decimal values printed beside them, such as
$0.870598$ for $0.870680\ldots$ in~\cite[eq.~(12)]{kamat1953}, are slightly off);
his tabulated $\beta_1$ and $\beta_2$~\cite[Table~1]{kamat1953} are the $\gamma_1^2$
and $\gamma_2+3$ of Table~\ref{tab:gamma-n}. What we add is the cumulant (connected-cluster) form, which recasts his exact moment formulas and, by the standard vanishing of joint cumulants of independent blocks~\cite{leonov_shiryaev1959}, has the same structure at every order, and a short proof of the closed form of the four-edge moment, which is due to Finch~\cite{finch2015} (Remark~\ref{rem:status}).

\begin{remark}[Numerical check of the constants]\label{rem:status}
Every constant of Table~\ref{tab:constants}, including~\eqref{eq:EZ1234} and each of the
six terms in the proof of Lemma~\ref{lem:E1234}, agrees with an independent quadrature
in the representation~\eqref{eq:S-msd} to at least $29$ digits (one- and
two-dimensional integrals after the outer $W$'s are integrated in closed form). For
general folded-normal moments see also~\cite{kanrobotti2017}; the reduction used here
is Gaussian integration by parts (Stein's identity), which is closely related to
Price's theorem~\cite{price1958}. Kamat~\cite[eqs.~(5)--(8)]{kamat1953} reduced
$\E|Z_1Z_2Z_3Z_4|$ to elementary terms and the integral
$I=\int_0^{1/2}\arcsin\sqrt{(1-x^2)/(3-4x^2)}\,dx/\sqrt{1-x^2}$, which he evaluated by
quadrature as $0.328988$; his value $2.215484$ corresponds to $4\E|Z_1Z_2Z_3Z_4|=2.2154831002\ldots$ in our normalisation (correct to six figures). Comparison with~\eqref{eq:EZ1234} through his reduction gives $I=\pi^2/30$, which we also confirmed by direct quadrature of $I$ to $40$ digits. The closed form of the four-edge moment is due to Finch~\cite{finch2015}, who obtained it, as $\mu_{1,4}=4\E|Z_1Z_2Z_3Z_4|$ in arctangent form (equal to~\eqref{eq:EZ1234} to $40$ digits), from Nabeya's four-variate formula~\cite[eq.~(12)]{nabeya1961}, whose
one non-elementary ingredient, $\E\sgn(Z_1Z_2Z_3Z_4)$, equals $2/15$ here, since the
signs of $W_{j+1}-W_j$ are the ascent pattern of a uniformly random permutation of five
points.
\end{remark}

\begin{corollary}[Skewness, kurtosis and the gamma ratios]\label{cor:asymptotics}
Let $\sigma_\infty^2=2A+4B=\frac83+\frac{4\sqrt3-12}\pi=1.0522636141$. As $n\to\infty$,
\begin{align*}
	\gamma_1(S)&=\frac{2\sqrt2\,C}{\sigma_\infty^3\sqrt n}+O(n^{-3/2})
	=\frac{1.5218221562}{\sqrt n}+O(n^{-3/2}),\\
	\gamma_2(S)&=\frac{4C_4}{\sigma_\infty^4\,n}+O(n^{-2})
	=\frac{2.8676851947}{n}+O(n^{-2}).
\end{align*}
The gamma law matched to $\E S$ and $\Var S$ has
\[
	\gamma_{1,\Gamma}=\frac{2\sqrt{\Var S}}{\E S}\sim\frac{\sqrt\pi\,\sigma_\infty}{\sqrt n}
	=\frac{1.8181814100}{\sqrt n},\qquad
	\gamma_{2,\Gamma}=\frac{6\Var S}{(\E S)^2}\sim\frac{3\pi\sigma_\infty^2}{2n}
	=\frac{4.9586754598}{n},
\]
so
\begin{align*}
	R_3^\infty&:=\lim_{n\to\infty}\frac{\gamma_1(S)}{\gamma_{1,\Gamma}}
	=\frac{2\sqrt2\,C}{\sqrt\pi\,\sigma_\infty^4}=0.8370023738,\\
	R_4^\infty&:=\lim_{n\to\infty}\frac{\gamma_2(S)}{\gamma_{2,\Gamma}}
	=\frac{8C_4}{3\pi\sigma_\infty^6}=0.5783167739 .
\end{align*}
\end{corollary}

\begin{proof}
Divide~\eqref{eq:kappa3-limit} and~\eqref{eq:kappa4-limit} by the powers
of~\eqref{eq:kappa2-limit} and use $\sigma^2=2/n$ and $\E S\sim(2/\sqrt\pi)\sqrt n$.
\end{proof}

Throughout, $R_3=\gamma_1/\gamma_{1,\Gamma}$ and $R_4=\gamma_2/\gamma_{2,\Gamma}$ for
whatever law is under discussion; the subscript $N$ marks exact finite-$N$ values and
the superscript $\infty$ the limit $n\to\infty$.
Since all constants are closed, $\gamma_1(S)$ and $\gamma_2(S)$ are explicit
elementary functions of $n$; Table~\ref{tab:gamma-n} lists some values. The
moment-matched gamma law therefore overstates the skewness of the limit by the
factor $1/R_3^\infty\approx1.19$ and the excess kurtosis by $1/R_4^\infty\approx1.73$.
The deficit is already present for a single folded normal: a single edge
($n=2$, the half-normal) has $R_3=a_3c/(2A^2)=0.6587$. The positive dependence of
adjacent $|G_j|$ ($B>0$, and the positive mixed cumulant $c_{21}$) raises the ratio towards $0.8370$, and the triple term
$c_{111}<0$ is a small correction. For contrast, a weighted sum $\sum\lambda_i\chi^2_1$,
the limit type of the Cram\'er--von Mises statistics, has
$R_3=\sum\lambda_i^3\sum\lambda_i/(\sum\lambda_i^2)^2\ge1$ by the Cauchy--Schwarz
inequality: there the gamma law is \emph{less} skewed than the target.

\begin{table}[ht]
\centering\small
\caption{Standardized cumulants of the limit law $S$ (Theorem~\ref{thm:cumulants}),
the moment-matched gamma values and their ratios (the values of $\gamma_1^2$ and
$\gamma_2+3$ for selected $n$ with $3\le n\le50$ are in Kamat~\cite[Table~1]{kamat1953}).}
\label{tab:gamma-n}
\begin{tabular}{rcccccc}
\toprule
$n$ & $\gamma_1(S)$ & $\gamma_{1,\Gamma}$ & $R_3$ & $\gamma_2(S)$ & $\gamma_{2,\Gamma}$ & $R_4$\\
\midrule
2   & 0.99527 & 1.51102 & 0.6587 & 0.86918 & 3.42478 & 0.2538\\
3   & 1.01762 & 1.18205 & 0.8609 & 1.29623 & 2.09586 & 0.6185\\
5   & 0.74479 & 0.87323 & 0.8529 & 0.68572 & 1.14380 & 0.5995\\
10  & 0.50311 & 0.59555 & 0.8448 & 0.31351 & 0.53203 & 0.5893\\
20  & 0.34785 & 0.41371 & 0.8408 & 0.14987 & 0.25673 & 0.5838\\
50  & 0.21710 & 0.25892 & 0.8385 & 0.05837 & 0.10056 & 0.5805\\
100 & 0.15285 & 0.18245 & 0.8377 & 0.02893 & 0.04993 & 0.5794\\
$\infty$ & $1.5218/\sqrt n$ & $1.8182/\sqrt n$ & 0.8370 & $2.8677/n$ & $4.9587/n$ & 0.5783\\
\bottomrule
\end{tabular}
\end{table}

\begin{proposition}[Cyclic statistic: exact formulas]\label{prop:cyclic}
Let $\ell_2=\sigma_\infty^2$, $\ell_3=2^{3/2}C$ and $\ell_4=4C_4$. Under the uniform
null, $\E S_c=2\sqrt{n/\pi}$ for $n\ge2$ and
\[
	\kappa_m(S_c)=n^{1-m/2}\,\ell_m\qquad\text{for } n\ge m+1,\ m=2,3,4 .
\]
In particular $\Var S_c=\sigma_\infty^2$ for $n\ge3$,
$\gamma_1(S_c)=\ell_3\ell_2^{-3/2}n^{-1/2}=1.5218221562\ldots/\sqrt n$ for $n\ge4$ and $\gamma_2(S_c)=\ell_4\ell_2^{-2}n^{-1}=2.8676851947\ldots/n$ for $n\ge5$, exactly.
\end{proposition}

\begin{proof}
In the cyclic representation $G_j=n^{-1/2}(W_{j+1}-W_j)$ with $W_{n+1}=W_1$, a joint
cumulant of order $m$ vanishes unless its index set is an arc of the ring, by the
argument of Lemma~\ref{lem:clusters}. If $n\ge m+1$, every such arc has at most
$m\le n-1$ edges and $m+1\le n$ distinct vertices, so its joint law is that of the
same pattern on the path, and each of the $n$ rotations occurs once. Hence
$\kappa_m(S_c)=n\cdot\sigma^m\cdot(\text{sum of the per-edge constants})$, which is
$n\sigma^2(A+2B)$, $n\sigma^3C$ and $n\sigma^4C_4$ for $m=2,3,4$.
\end{proof}

The value $\Var S_c=\sigma_\infty^2$ is the limit $\alpha_c$
of~\cite[Theorem~3.1]{banic_elezovic_p5}, and $\E S_c$ is the leading term
of~\cite[Theorem~5.5]{banic_elezovic_p1}. At $n=m$ the arc consisting of all $n$
edges closes on itself and the formula changes; for $n=2$, $G_2=-G_1$ and
$S_c=2|G_1|$.

The same counting works for every order $m\ge2$ (for $m=1$ the centering of the $H_j$
removes the mean, and $\E S$ is given in Theorem~\ref{thm:cumulants}): with $a_{m,L}$ the sum of the joint
cumulants $\kappa(H_{i_1},\ldots,H_{i_m})$ over ordered tuples whose index set is
$\{1,\ldots,L\}$, one has $\kappa_m(S)=\sigma^m\sum_{L=1}^m(n-L)a_{m,L}$ for $n\ge m$ and
$\kappa_m(S_c)=n^{1-m/2}2^{m/2}\sum_{L=1}^ma_{m,L}$ for $n\ge m+1$; only the constants $a_{m,L}$ for $m\ge5$ are not known in elementary form.

\begin{remark}[A transfer operator]\label{rem:transfer}
By~\eqref{eq:S-msd}, $\E e^{t\sqrt n\,S_c}=\operatorname{Tr}K_t^n$ for the symmetric
kernel $K_t(x,y)=\sqrt{\varphi(x)}\,e^{t|x-y|}\sqrt{\varphi(y)}$ on $\mathbb R$, and
$\E e^{t\sqrt n\,S}=\langle\sqrt\varphi,K_t^{n-1}\sqrt\varphi\rangle$. The per-edge
cumulants $\ell_m$ are the derivatives at $t=0$ of the logarithm of the top eigenvalue
of $K_t$. This gives a route to the higher cumulants
(Section~\ref{sec:open}).
\end{remark}

\begin{proposition}[Gaussian extreme tail]\label{prop:tail}
Let $\bar\Phi=1-\Phi$. For $n\ge2$, as $t\to\infty$,
\[
	\Pr(S>t)\sim2\,\bar\Phi\bigl(t/\sigma_*\bigr),\qquad \sigma_*^2=4-\frac6n .
\]
For the cyclic statistic, $\Pr(S_c>t)\sim2\bar\Phi(t/2)$ if $n\ge2$ is even, and
$\Pr(S_c>t)\sim2n\,\bar\Phi\bigl(t/\sqrt{4-4/n}\bigr)$ if $n\ge3$ is odd.
\end{proposition}

\begin{proof}
$S=\max_{\varepsilon\in\{\pm1\}^{n-1}}\sum_j\varepsilon_jG_j$, and
$\Var\sum_j\varepsilon_jG_j=\frac2n\{(n-1)-\sum_j\varepsilon_j\varepsilon_{j+1}\}$. This is
maximal, equal to $4-6/n$, exactly for the two alternating sign vectors $\pm\varepsilon^*$,
and every other $\varepsilon$ has a strictly smaller variance. Hence
$2\bar\Phi(t/\sigma_*)=\Pr(|\langle\varepsilon^*,G\rangle|>t)\le\Pr(S>t)
\le2\bar\Phi(t/\sigma_*)+\sum_{\varepsilon\ne\pm\varepsilon^*}\bar\Phi(t/\sigma_\varepsilon)$,
and the sum is $o(\bar\Phi(t/\sigma_*))$. On the ring the variance is
$\frac2n\{n-\sum_j\varepsilon_j\varepsilon_{j+1}\}$; for even $n$ the alternating vectors give
the unique maximum $4$, and for odd $n$ at least one bond is not alternating, so the
maximum is $4-4/n$, attained by $2n$ sign vectors, which form $n$ antipodal pairs
$\pm\varepsilon$. Since the covariance of the cyclic edge vector has kernel spanned by
$\mathbf 1$, two of these forms coincide only if the sign vectors are equal, and are
opposite only if the sign vectors are opposite (a constant sign vector has variance $0$).
For an antipodal pair, $Y=-X$, we have $\Pr(X>t,Y>t)=0$ for $t>0$. Every other pair
$X,Y$ has correlation $-1<\rho<1$, and
$\Pr(X>t,Y>t)\le\Pr(X+Y>2t)=\bar\Phi(t/\tau)$ with
$0<\tau^2=\Var\frac{X+Y}2=\frac{1+\rho}2\,(4-\frac4n)<4-\frac4n$, so that
$\Pr(X>t,Y>t)=o\bigl(\bar\Phi(t/\sqrt{4-4/n})\bigr)$. (The comparison must be with the
maximal variance $4-4/n$, not with $\sigma_*^2$: for the most correlated pair $\tau^2$
equals $4-6/n=\sigma_*^2$.) By inclusion--exclusion and the
same upper bound as above, $\Pr(S_c>t)\sim2n\,\bar\Phi(t/\sqrt{4-4/n})$. For $n=2$, $S_c=2|G_1|$ with $\Var G_1=1$.
\end{proof}

Non-asymptotically, since $S$ is the maximum of the finitely many Gaussian forms
$\langle\varepsilon,G\rangle$, of largest variance $\sigma_*^2$, the Borell--TIS
inequality~\cite{borell1975,tsirelson_ibragimov_sudakov1976} gives $\Pr(S-\E S\ge t)\le\exp\{-t^2/(2\sigma_*^2)\}$ for
every $t>0$.

The variance proxy $\sigma_*^2\to4$ is about four times $\Var S\to\sigma_\infty^2$, so
the tail is much heavier than that of the normal law with the same variance, but it is
still Gaussian. A gamma tail decays only exponentially. For fixed $n$ the
moment-matched gamma quantile $q^\Gamma_{1-\alpha}\sim\theta\log(1/\alpha)$, with the
gamma scale $\theta=\Var S/\E S$, therefore
exceeds the true quantile, which grows like $\sigma_*\sqrt{2\log(1/\alpha)}$, by an
amount that tends to infinity as $\alpha\to0$. This is a structural reason why, for fixed $n$, the overshoot of the gamma law grows without bound as $\alpha\to0$; at the levels of Section~\ref{sec:tails} the overshoot is already present through the smaller skewness and kurtosis ($R_3,R_4<1$).

\paragraph{Large $n$ and the Edgeworth correction.}
By~\eqref{eq:S-msd} $S$ is a sum of $n-1$ one-dependent terms, so
$(S-\E S)/\sqrt{\Var S}\Rightarrow\mathcal N(0,1)$ as $n\to\infty$~\cite{hoeffding_robbins1948},
with $\gamma_1(S)=O(n^{-1/2})$ the leading departure from normality. This expansion
concerns the second limit $n\to\infty$ of the already formed limit law $S$; it is not
an approximation theorem for $d$ at finite $N$. With $z=(x-\E S)/\sqrt{\Var S}$ the
formal second-order Edgeworth expansion (of the type justified for weakly dependent
sums in~\cite{gotze_hipp1983}, whose conditions we do not verify here) reads
\begin{equation}\label{eq:edgeworth-limit-law}
	\Pr(S\le x)
	\approx
	\Phi(z)-\varphi(z)\Bigl\{\frac{\gamma_1(S)}6\,(z^2-1)
	+\frac{\gamma_2(S)}{24}(z^3-3z)
	+\frac{\gamma_1(S)^2}{72}(z^5-10z^3+15z)\Bigr\},
\end{equation}
with $\gamma_1(S)$ and $\gamma_2(S)$ from Theorem~\ref{thm:cumulants}.

\subsection{A general null: the profile}
	\label{ssec:profile}

Let $\mathbf r$ be a null with all $r_i>0$, and let $Q_{\mathbf r}=\sum_i|G_i|$,
$G_i=G_i^{(\mathbf r)}$, be the limit of $d_{\mathbf r}/\sqrt N$
(Theorem~\ref{thm:limit}(b)). Write $s_i^2=\Var G_i=r_i+r_{i+1}-\Delta_i^2$ and
$\rho_{ij}=\Corr(G_i,G_j)$. The edge covariance need not be tridiagonal, but it is always a tridiagonal covariance minus a rank-one term, which comes from one common factor.

\begin{proposition}[Moving average corrected by one common factor]\label{prop:rank-one}
Let $V_1,\ldots,V_n$ be independent, $V_i\sim\mathcal N(0,r_i)$, and $T=V_1+\cdots+V_n$.
Then
\[
	\bigl(G_1^{(\mathbf r)},\ldots,G_{n-1}^{(\mathbf r)}\bigr)
	\overset{d}{=}
	\bigl(V_2-V_1-\Delta_1T,\ \ldots,\ V_n-V_{n-1}-\Delta_{n-1}T\bigr),
\]
and $T\sim\mathcal N(0,1)$ is independent of the vector on the right-hand side.
\end{proposition}

\begin{proof}
The vector $Y_i=V_i-r_iT$ is centered Gaussian with
$\E Y_iY_j=r_i\delta_{ij}-2r_ir_j+r_ir_j\sum_kr_k=r_i\delta_{ij}-r_ir_j$, the normalized
multinomial covariance $\operatorname{diag}\mathbf r-\mathbf r\mathbf r^{\!\top}$. By the
proof of Theorem~\ref{thm:limit}(a), $\mathbf G^{(\mathbf r)}$ is distributed as
$(Y_{i+1}-Y_i)_i=(V_{i+1}-V_i-\Delta_iT)_i$. Finally $\Var T=\sum_ir_i=1$, and $\Cov(Y_i,T)=r_i-r_i\Var T=0$, so the Gaussian vector $(Y_i)$, and with it the right-hand side, is independent of $T$.
\end{proof}

Under the uniform null $\Delta_i=0$ and $V_i=n^{-1/2}W_i$, which
is~\eqref{eq:S-msd}. In general the one-dependent part $V_{i+1}-V_i$ has the
tridiagonal covariance $C_0$ (diagonal $r_i+r_{i+1}$, neighbours $-r_{i+1}$); since
$\Cov(V_{i+1}-V_i,T)=\Delta_i$, the correction $-\Delta_iT$ turns it into
$C_0-\Delta\Delta^{\!\top}$, which gives the long-range covariance $-\Delta_i\Delta_j$
of Theorem~\ref{thm:limit}(a).

The assumption that all $r_i>0$ is a genuine restriction. A cell with $r_i=0$ (a
structural zero) is empty under the null, but it cannot simply be deleted: the
statistic is formed on the original ordered cells, and deleting the cell makes its two
neighbours adjacent and changes both $d_{\mathbf r}$ and its law. For
$\mathbf r=(0.2,0,0.3,0.5)$, $N=10$ and $X=(3,0,4,3)$ the statistic is $5$ on the
original cells but $3$ after deletion, and the null means are $4.2788$ and $3.9352$.
In the limit such a cell has residual $0$, so an edge joining it to a positive cell
contributes the absolute residual of that cell, and the covariance of
$\mathbf G^{(\mathbf r)}$ is singular. Profiles with structural zeros are not treated
here.

For the moments of $Q_{\mathbf r}$ we need the absolute moments of a Gaussian vector
with an arbitrary correlation matrix. Put
\[
	m_2(\rho)=\frac2\pi\bigl(\sqrt{1-\rho^2}+\rho\arcsin\rho\bigr),
\]
and for a positive definite correlation matrix $(\rho_{ab})_{a,b=1}^3$ with
off-diagonal entries $\rho_{12},\rho_{13},\rho_{23}$ we write
\[
	m_3(\rho_{12},\rho_{13},\rho_{23})=\Bigl(\frac2\pi\Bigr)^{3/2}\Bigl[\sqrt{\det(\rho_{ab})}
	+\sum(\rho_{ab}+\rho_{ac}\rho_{bc})
	\arcsin\frac{\rho_{ab}-\rho_{ac}\rho_{bc}}{\sqrt{(1-\rho_{ac}^2)(1-\rho_{bc}^2)}}\Bigr],
\]
the sum running over the three pairs $\{a,b\}$, with $c$ the remaining index. For unit
normal variables, $\E|Z_1Z_2|=m_2(\rho_{12})$ (Nabeya~\cite{nabeya1951}) and
$\E|Z_1Z_2Z_3|=m_3$ (Nabeya~\cite{nabeya1952}); at the correlations of three consecutive
uniform edges $m_3$ is~\eqref{eq:EZ123}.

\begin{theorem}[The third cumulant for every positive null]\label{thm:kappa3-profile}
With $c=\sqrt{2/\pi}$,
\begin{align*}
	\E Q_{\mathbf r}&=c\sum_is_i,\qquad
	\E Q_{\mathbf r}^2=\sum_is_i^2+2\sum_{i<j}s_is_j\,m_2(\rho_{ij}),\\
	\E Q_{\mathbf r}^3&=2c\sum_is_i^3+3c\sum_{i\ne j}s_i^2s_j(1+\rho_{ij}^2)
	+6\sum_{i<j<k}s_is_js_k\,m_3(\rho_{ij},\rho_{ik},\rho_{jk}),
\end{align*}
all indices running over $1,\ldots,n-1$, so that
$\kappa_2(Q_{\mathbf r})=\E Q_{\mathbf r}^2-(\E Q_{\mathbf r})^2$ and
$\kappa_3(Q_{\mathbf r})=\E Q_{\mathbf r}^3-3\E Q_{\mathbf r}\,\E Q_{\mathbf r}^2
+2(\E Q_{\mathbf r})^3$ are elementary functions of $\mathbf r$.
\end{theorem}

\begin{proof}
Expand $Q_{\mathbf r}^3=\sum_{i,j,k}|G_iG_jG_k|$ and group the triples by the number
of distinct indices. For one index, $\E|G_i|^3=2c\,s_i^3$. For two, write
$G_i/s_i=\rho_{ij}G_j/s_j+\sqrt{1-\rho_{ij}^2}\,\varepsilon$ with $\varepsilon$ independent
of $G_j$; then $\E G_i^2|G_j|=s_i^2s_j\{\rho_{ij}^2\E|Z|^3+(1-\rho_{ij}^2)\E|Z|\}
=c\,s_i^2s_j(1+\rho_{ij}^2)$, and each unordered pair $\{i,j\}$ with one index doubled
occurs in $3$ ordered positions. Three distinct indices occur in $3!=6$ orders and give
Nabeya's trivariate moment. The second moment is the same computation with Nabeya's
bivariate moment. The covariance of $\mathbf G^{(\mathbf r)}$ is nonsingular (the edges
form a basis of the hyperplane $\sum_ix_i=0$), so all correlations, and the partial
correlations under the arcsines, lie in $(-1,1)$.
\end{proof}

The variance $\kappa_2(Q_{\mathbf r})$ is the leading variance coefficient of
$d_{\mathbf r}$ found in~\cite[Corollary~3.5]{banic_elezovic_p4} (and expanded in~\cite[Theorem~3.3]{banic_elezovic_p4} for $n\ge5$ and in~\cite[Corollary~6.4]{banic_elezovic_p4} for $n\ge3$);
the new part of Theorem~\ref{thm:kappa3-profile} is $\kappa_3$. Under the uniform null
the formulas reduce to~\eqref{eq:kappa2-limit} and~\eqref{eq:kappa3-limit} (we checked
the agreement to $33$ digits for $3\le n\le7$). Off the uniform null, for a generic profile, non-adjacent triples of edges also contribute, through the common factor of Proposition~\ref{prop:rank-one} (for special profiles such as $(1,2,2,2,2)/9$ none do), so
$\kappa_3(Q_{\mathbf r})$ is a sum of $O(n^3)$ elementary terms rather than an affine
function of $n$. For $n\le4$ there are at most three edges, and
$\kappa_4(Q_{\mathbf r})$ is elementary as well (the first four moments of a weighted
sum of two or three correlated folded normals are
in~\cite[eqs.~(50), (51)]{kamat1953a}). For $n\ge5$ the fourth cumulant needs
four-variate absolute moments for a general correlation matrix. These are available as
elementary terms plus one orthant-type integral
(Nabeya~\cite[eq.~(12)]{nabeya1961}; see also the formulas of~\cite{kanrobotti2017},
which involve quadrivariate normal orthant probabilities, and~\cite{benko2025}, where
the moment generating function of a sum of dependent folded normals such as
$Q_{\mathbf r}$ is expressed through multivariate normal distribution functions), so
$\kappa_4(Q_{\mathbf r})$ is computable for every $\mathbf r$, but we do not have an
elementary form comparable to Table~\ref{tab:constants} (Section~\ref{sec:open}). By an
elementary form we mean a finite expression in the profile entries $r_i$ (or in the
correlations $\rho_{ij}$) built from arithmetic operations, square roots, $\pi$ and
inverse trigonometric functions, with no unevaluated multivariate normal probability
or integral; for the uniform null, or another profile with algebraic entries, its value
is then an expression in $\pi$, square roots and inverse trigonometric values of
algebraic numbers, as in Table~\ref{tab:constants}. Normal orthant probabilities of
dimension four or more are not counted. For large $n$ and a smooth profile, however, the common factor becomes negligible, and $\kappa_3$ and $\kappa_4$ follow at leading order from the uniform cluster constants weighted by the profile.

\begin{theorem}[The profile law]\label{thm:profile-law}
Let $f\in C^1([0,1])$ be positive, let $\mathbf r=\mathbf r^{(n)}$ be the triangular
array $r_i=f(i/n)/\sum_{k=1}^nf(k/n)$, $i=1,\ldots,n$, and write
$I_a=\int_0^1f^a\big/\bigl(\int_0^1f\bigr)^a$. As $n\to\infty$,
$\kappa_2(Q_{\mathbf r})\to\sigma_\infty^2$, and
\begin{align*}
	\sqrt n\,\gamma_1(Q_{\mathbf r})&\to\frac{2\sqrt2\,C}{\sigma_\infty^3}\,I_{3/2}
	=1.5218221562\,I_{3/2},\\
	n\,\gamma_2(Q_{\mathbf r})&\to\frac{4C_4}{\sigma_\infty^4}\,I_2
	=2.8676851947\,I_2 .
\end{align*}
For the ratios $R_3=\gamma_1/\gamma_{1,\Gamma}$ and $R_4=\gamma_2/\gamma_{2,\Gamma}$ to the
moment-matched gamma law (Corollary~\ref{cor:asymptotics}),
\[
	R_3\to R_3^\infty\,I_{3/2}\,I_{1/2},\qquad R_4\to R_4^\infty\,I_2\,I_{1/2}^2 .
\]
\end{theorem}

\begin{proof}
Constants $K$ below depend on $f$ and on the order of the norms only. Since $f$ is
positive and $C^1$, $r_i\asymp1/n$, $|r_{i+1}/r_i-1|\le K/n$ and $|\Delta_i|\le K/n^2$.

\emph{The one-dependent part.} By Proposition~\ref{prop:rank-one},
$G_i=G_i^0-\Delta_iT$ with $G_i^0=V_{i+1}-V_i$. Let $Q^0=\sum_i|G_i^0|$. Each $|G_i^0|$ is a
function of $(V_i,V_{i+1})$, so, as in Lemma~\ref{lem:clusters}, a joint cumulant of
$|G^0_{i_1}|,\ldots,|G^0_{i_m}|$ vanishes unless $\{i_1,\ldots,i_m\}$ is an interval. If
the interval starts at $i$ and has $\ell\le m$ edges, the joint cumulant is a function
$\Psi(r_i,\ldots,r_{i+\ell})$. With $V_j=\sqrt{r_j}\,U_j$, $U_j$ i.i.d.\ standard normal,
$\Psi$ is positively homogeneous of degree $m/2$, and it is locally Lipschitz on
$(0,\infty)^{\ell+1}$: a joint moment $\E\prod_t|a_tU_{t+1}-b_tU_t|$ is Lipschitz in the
coefficients by $\bigl||x|-|y|\bigr|\le|x-y|$ and H\"older's inequality, and a cumulant is a
polynomial in moments; we use the Lipschitz constant on the compact set where all
ratios $r_{i+j}/r_i$ lie in $[\frac12,2]$, which contains every cluster for large $n$. Hence
$\Psi(r_i,\ldots,r_{i+\ell})=r_i^{m/2}\Psi(1+O(n^{-1}),\ldots)=r_i^{m/2}\{\Psi(1,\ldots,1)+O(n^{-1})\}$,
and $\Psi(1,\ldots,1)$ is the uniform cluster cumulant of the edges $U_{j+1}-U_j$, of
variance $2$. Grouping the clusters by their first edge as in the proof of
Theorem~\ref{thm:cumulants}, the clusters starting at $i\le n-m$ contribute
$r_i^{m/2}\{\ell_m+O(n^{-1})\}$, with the per-edge constants $\ell_2=2(A+2B)=\sigma_\infty^2$,
$\ell_3=2^{3/2}C$, $\ell_4=4C_4$ of Proposition~\ref{prop:cyclic}, and the $O(1)$
clusters near the right end contribute $O(n^{-m/2})$. With the Riemann sum
$\sum_{i=1}^nr_i^{m/2}=n^{1-m/2}\{I_{m/2}+O(n^{-1})\}$ this gives
\begin{equation}\label{eq:profile-Q0}
	\kappa_m(Q^0)=\ell_m\,n^{1-m/2}\,I_{m/2}+O(n^{-m/2}),\qquad m=2,3,4 .
\end{equation}

\emph{The common factor.} Let $E=Q_{\mathbf r}-Q^0=\sum_i(|G_i^0-\Delta_iT|-|G_i^0|)$. For
real $y,h$ one has $|y-h|-|y|=-h\sgn y$ if $|y|>|h|$, and
$\bigl||y-h|-|y|+h\sgn y\bigr|\le2|h|$ always; hence
$E=-T\sum_i\Delta_i\sgn G_i^0+R$ with $|R|\le2|T|\sum_i|\Delta_i|\mathbf 1\{|G_i^0|\le|\Delta_iT|\}$.
The signs $\sgn G_i^0$ are centered and one-dependent, so, splitting into odd and even
$i$ and applying the Marcinkiewicz--Zygmund inequality to each half,
$\|\sum_i\Delta_i\sgn G_i^0\|_p\le K(\sum_i\Delta_i^2)^{1/2}\le Kn^{-3/2}$, and by
H\"older's inequality the first term of $E$ has $L^p$ norm at most $Kn^{-3/2}$. The
density of $G_i^0$ is at most $(2\pi(r_i+r_{i+1}))^{-1/2}\le K\sqrt n$, so
$\Pr(|G_i^0|\le|\Delta_iT|)\le\Pr(|T|>\sqrt n)+\Pr(|G_i^0|\le\sqrt n|\Delta_i|)\le K/n$,
and
$\|T\Delta_i\mathbf 1\{\cdot\}\|_p\le|\Delta_i|\,\|T\|_{2p}\Pr(\cdot)^{1/(2p)}\le Kn^{-2-1/(2p)}$.
Therefore $\|E\|_2\le Kn^{-5/4}$ and $\|E\|_p\le K_pn^{-1}$ for every $p$.

\emph{Comparison.} Let $Y=Q^0-\E Q^0$ and $\tilde E=E-\E E$. The variable $Y$ is a sum of
one-dependent centered terms with $L^p$ norms $O(n^{-1/2})$, so $\|Y\|_p\le K_p$ by
Rosenthal's inequality applied to the odd and the even terms. Expanding
$\kappa_m(Q_{\mathbf r})=\kappa_m(Y+\tilde E)$ for $m=2,3,4$ in powers of $\tilde E$, every
term containing $\tilde E$ is bounded by H\"older's inequality by a product of norms of
$Y$ and either one factor $\|\tilde E\|_2$ or at least two factors $\|\tilde E\|_q$ with
$q\le8$; hence
$\kappa_m(Q_{\mathbf r})=\kappa_m(Q^0)+O(n^{-5/4})$. By~\eqref{eq:profile-Q0},
$\kappa_2\to\ell_2=\sigma_\infty^2$, $\sqrt n\,\kappa_3\to2^{3/2}C\,I_{3/2}$ and
$n\kappa_4\to4C_4\,I_2$, which gives $\gamma_1$ and $\gamma_2$; the numerical constants
are those of Corollary~\ref{cor:asymptotics}. (Numerically the bound $O(n^{-5/4})$ is very
conservative: $\kappa_2(Q_{\mathbf r})-\kappa_2(Q^0)$ is of order $n^{-3}$ for the profiles of
Remark~\ref{rem:profile-check}.) Finally $\E Q_{\mathbf r}=c\sum_is_i$
exactly, with $s_i=\sqrt{2r_i}\,\{1+O(n^{-1})\}$ and
$\sum_i\sqrt{r_i}=\sqrt n\,\{I_{1/2}+O(n^{-1})\}$, so
$\sqrt n\,\gamma_{1,\Gamma}\to\sqrt\pi\,\sigma_\infty/I_{1/2}$ and
$n\gamma_{2,\Gamma}\to\frac{3\pi}2\sigma_\infty^2/I_{1/2}^2$; dividing gives $R_3$ and $R_4$.
\end{proof}

The limit $\kappa_2\to\sigma_\infty^2$, independent of $f$, is the smooth-profile
universality of~\cite[Proposition~4.4]{banic_elezovic_p4}. Theorem~\ref{thm:profile-law}
shows that the universality stops at the variance: the skewness and the excess kurtosis carry
the profile at leading order, through $I_{3/2}$ and $I_2$.

\begin{corollary}[The uniform profile minimises the smooth-profile constants]\label{cor:uniform-min}
Under the assumptions of Theorem~\ref{thm:profile-law}, the limits of
$\sqrt n\,\gamma_1$, $n\gamma_2$, $R_3$ and $R_4$ are at least their uniform values
$1.5218$, $2.8677$, $R_3^\infty=0.8370$ and $R_4^\infty=0.5783$, with equality if and only
if $f$ is constant.
\end{corollary}

\begin{proof}
By Jensen's inequality $I_{3/2}\ge1$ and $I_2\ge1$. By the Cauchy--Schwarz inequality
$(\int f)^2=(\int f^{3/4}f^{1/4})^2\le\int f^{3/2}\int f^{1/2}$, that is,
$I_{3/2}I_{1/2}\ge1$; and by H\"older's inequality with exponents $3$ and $\frac32$,
$\int f=\int(f^2)^{1/3}(f^{1/2})^{2/3}\le(\int f^2)^{1/3}(\int f^{1/2})^{2/3}$, that is,
$I_2I_{1/2}^2\ge1$. In each case equality holds only for constant $f$.
\end{proof}

So for each fixed smooth non-uniform $f$ and all sufficiently large $n$ the limit law is more skewed and more kurtotic than under the uniform null, and its ratio $R_3$ moves towards the gamma value $1$, and for strong profiles beyond it (once $R_3^\infty I_{3/2}I_{1/2}>1$, that is,
$I_{3/2}I_{1/2}>1.195$). The effect is of first order in $n^{-1/2}$ for $\gamma_1$ (in $n^{-1}$ for $\gamma_2$), but its size $I_{3/2}-1$ is small for mild profiles:
$1.4\%$ for $f(x)=1+x$, $4.7\%$ for $1+\frac12\cos2\pi x$, and $11.4\%$ for $e^{2x}$.
The convergence is not uniform in $f$: for $f_k(x)=1+0.9\cos2\pi kx$, whose $I_{3/2}$ does not depend on $k\ge1$, the choice $k=n/2$ (weights alternating $1.9:0.1$) gives, with the exact $\kappa_3$ of Theorem~\ref{thm:kappa3-profile}, $\sqrt n\,\gamma_1(Q_{\mathbf r})=1.499$ at $n=400$, below the uniform value $1.523$ at the same $n$.

\begin{remark}[Numerical check]\label{rem:profile-check}
With the exact $\kappa_3$ of Theorem~\ref{thm:kappa3-profile} (full covariance), the
values $a_n=\sqrt n\,\gamma_1(Q_{\mathbf r})$ at $n=100,200,400$ and the
extrapolation $2a_{400}-a_{200}$ in $1/n$ are given in Table~\ref{tab:profile-check}.
The ratio $R_3$ agrees with $R_3^\infty I_{3/2}I_{1/2}$ to within $7\cdot10^{-6}$ after
the same extrapolation. For $\gamma_2$ a Monte Carlo sample of $8\cdot10^6$ draws at
$n=60$ through Proposition~\ref{prop:rank-one}, with common random numbers, gives the
ratios $n\gamma_2(Q_{\mathbf r})/n\gamma_2(S)=1.039\pm0.009$, $1.114\pm0.022$ and
$1.304\pm0.025$ (batch-means standard errors from $16$ batches of $5\cdot10^5$
draws) for the last three profiles, against $I_2=1.037$, $1.125$ and $1.313$.
\end{remark}

\begin{table}[ht]
\centering\small
\caption{Profile skewness at large $n$: $a_n=\sqrt n\,\gamma_1(Q_{\mathbf r})$ from the
exact $\kappa_3$ of Theorem~\ref{thm:kappa3-profile} (full covariance, including the
common factor) for the profile $r_i\propto f(i/n)$, the extrapolation
$2a_{400}-a_{200}$ in $1/n$, and the limit $1.5218221562\,I_{3/2}$ of
Theorem~\ref{thm:profile-law} (Remark~\ref{rem:profile-check}).}
\label{tab:profile-check}
\begin{tabular}{lccccc}
\toprule
$f$ & $n=100$ & $n=200$ & $n=400$ & extrapolated & $1.5218221562\,I_{3/2}$\\
\midrule
$1$ & 1.528458 & 1.525132 & 1.523475 & 1.521818 & 1.521822\\
$1+x$ & 1.548987 & 1.546009 & 1.544527 & 1.543044 & 1.543049\\
$1+\frac12\cos2\pi x$ & 1.598104 & 1.596110 & 1.595088 & 1.594066 & 1.594050\\
$e^{2x}$ & 1.696417 & 1.696028 & 1.695838 & 1.695649 & 1.695653\\
\bottomrule
\end{tabular}
\end{table}

At fixed small $n$ the profile acts differently, and the uniform null does not minimise the skewness. For $n=3$ everything is explicit.

\begin{proposition}[Three cells]\label{prop:n3}
Let $n=3$ and all $r_i>0$.
\begin{enumerate}
\item[(a)] For every outcome,
$d_{\mathbf r}=\max\bigl(|X_3-X_1-N(r_3-r_1)|,\,3|X_2-Nr_2|\bigr)$; in particular
$d=\max(|X_3-X_1|,|3X_2-N|)$.
\item[(b)] $Q_{\mathbf r}=\max(|\xi|,|\eta|)$, where $(\xi,\eta)$ is centered Gaussian with
$\Var\xi=r_1+r_3-(r_3-r_1)^2$, $\Var\eta=9r_2(1-r_2)$ and $\Cov(\xi,\eta)=3r_2(r_1-r_3)$; thus
$\Pr(Q_{\mathbf r}\le x)=\Pr(|\xi|\le x,\,|\eta|\le x)$.
\item[(c)] If $r_1=r_3$, then $\xi$ and $\eta$ are independent and
\[
	\Pr(Q_{\mathbf r}\le x)=\operatorname{erf}\Bigl(\frac{x}{\sqrt{2\Var\xi}}\Bigr)
	\operatorname{erf}\Bigl(\frac{x}{\sqrt{2\Var\eta}}\Bigr),\qquad x\ge0;
\]
under the uniform null $\Pr(S\le x)=\operatorname{erf}(\sqrt3\,x/2)\operatorname{erf}(x/2)$ for $x\ge0$.
\end{enumerate}
\end{proposition}

\begin{proof}
(a) Apply $|a|+|b|=\max(|a+b|,|a-b|)$ to $a=\tilde D_1$, $b=\tilde D_2$: $a+b=X_3-X_1-N(r_3-r_1)$,
and $a-b=2X_2-X_1-X_3-N(2r_2-r_1-r_3)=3(X_2-Nr_2)$ because $X_1+X_2+X_3=N$ and
$r_1+r_2+r_3=1$. (b) Divide by $\sqrt N$ and apply the central limit theorem and the
continuous mapping theorem to the linear images $X_3-X_1$ and $3X_2$ of the multinomial
vector; their normalized covariances are the displayed ones. (c) is immediate from
$\Cov(\xi,\eta)=0$.
\end{proof}

Part~(b) is the case of the representation of $\Pr\{|x_1|+c|x_2|<d\}$ by bivariate
normal integrals in~\cite[\S4.2]{kamat1953a}; under the uniform null it
is~\cite[\S2.4]{kamat1953}, and~(c) is the simplification noted in
Remark~\ref{rem:density}.

By~(a), the exact law of $d_{\mathbf r}$ for $n=3$ needs only binomial distribution
functions: given $X_2=j$, $X_1\sim\mathrm{Bin}(N-j,r_1/(r_1+r_3))$ and
$X_3-X_1=N-j-2X_1$. By~(b) the limit law depends on $\mathbf r$ only through the covariance of $(\xi,\eta)$, and its standardized cumulants only through the correlation of $\xi$ and $\eta$ and the ratio of their variances. The moments of
$\max(|\xi|,|\eta|)$ reduce in polar coordinates to one-dimensional integrals; they agree with
Theorem~\ref{thm:kappa3-profile} to $30$ digits and give the limit values of
Table~\ref{tab:n3-limit}. The first row is the uniform null, and the last row,
$(4,1,4)$, is the profile with uncorrelated edges of equal variance, where
$Q_{\mathbf r}$ is a scaled sum of two independent half-normal variables (Numerical
observation~\ref{obs:n3-range}).

\begin{table}[ht]
\centering\small
\caption{Limit values at $n=3$ for five profiles $\mathbf r$: standardized cumulants of
$Q_{\mathbf r}$ and the ratios $R_3$, $R_4$ to the moment-matched gamma law, from the
one-dimensional integrals of $\max(|\xi|,|\eta|)$ in polar coordinates
(Proposition~\ref{prop:n3}(b)).}
\label{tab:n3-limit}
\begin{tabular}{lcccc}
\toprule
$\mathbf r\propto$ & $\gamma_1(Q_{\mathbf r})$ & $\gamma_2(Q_{\mathbf r})$ & $R_3$ & $R_4$\\
\midrule
$(1,1,1)$ & 1.017616 & 1.296227 & 0.8609 & 0.6185\\
$(1,2,1)$ & 1.101564 & 1.407483 & 0.8813 & 0.6006\\
$(1,3,1)$ & 1.120388 & 1.401374 & 0.8759 & 0.5710\\
$(1,2,3)$ & 1.074517 & 1.371146 & 0.8746 & 0.6057\\
$(4,1,4)$ & 0.703763 & 0.434589 & 0.6587 & 0.2538\\
\bottomrule
\end{tabular}
\end{table}

\begin{observation}[The range at $n=3$]\label{obs:n3-range}
Over the profiles with $n=3$ and all $r_i>0$, a numerical search found
$\gamma_1(Q_{\mathbf r})$ in $[0.703763,\,1.132507]$. The search evaluated the closed
form of Theorem~\ref{thm:kappa3-profile} on the product grid $(r_1,r_2)\in E\times E$,
$r_1+r_2<1-10^{-7}$, where $E$ consists of $60$ logarithmically spaced points in
$[10^{-7},10^{-2}]$ and $2000$ equally spaced points in $[0.01,0.99]$
($2\,242\,598$ evaluations, $2\,238\,481$ distinct points of the simplex, since $0.01$ belongs to both parts), started Nelder--Mead in the coordinates
$(r_1,r_2)$ from the grid minimiser and maximiser (tolerances $10^{-12}$ in the argument
and $10^{-15}$ in the value), and was repeated, with tolerances $10^{-10}$ and $10^{-14}$, from $200$ random starts in logarithmic
coordinates $r_i\propto e^{u_i}$, $u_3=0$, with $(u_1,u_2)$ started uniformly in
$[-12,12]^2$ and confined to $[-16,16]^2$. The values above are reproduced by $33$ of
the $200$ minimisations and $198$ of the $200$ maximisations; the other runs stop in the
boundary layer $\min_ir_i<10^{-5}$ at larger, respectively smaller, values. As one
$r_i\to0$ alone, $\gamma_1$ tends to $0.995272$, the value for two cells; at the vertices
the limit depends on the direction of approach, with values between $0.732842$ and $1.132256$. The
smallest value found is at $\mathbf r=(4,1,4)/9$, where the two edges are
uncorrelated with equal variances, so that $Q_{\mathbf r}$ is $\sqrt{4/9}$ times a sum of
two independent half-normal variables and $\gamma_1=0.99527/\sqrt2$. The largest value
found is attained at two symmetric profiles, $\mathbf r\approx(0.0738,0.8525,0.0738)$ and
$\mathbf r\approx(0.4928,0.0145,0.4928)$, with equal edge variances and edge correlation
$\mp0.769$ (on the line $r_1=r_3=(1-b)/2$ the edge correlation is $(1-9b)/(1+9b)$, and
$\gamma_1$ is even in it). These are candidate extrema of a numerical search; the
endpoints of the range are not certified.
\end{observation}

So at $n=3$ the profile moves the skewness by about $-31\%$ and $+11\%$ around the uniform value (the true range is at least as wide as the one found), in both directions, and the gamma ratio $R_3$ can fall below or rise above its
uniform value; Corollary~\ref{cor:uniform-min} is a large-$n$ statement.

\section{Finite-\(N\) approach to the limiting cumulants}
	\label{sec:finite-N}

Throughout this section the null is uniform, $n$ is fixed and $N\to\infty$. By
Theorem~\ref{thm:limit}(b) the standardized cumulants $\gamma_m(N,n)$ of $d$ converge to
$\gamma_m^\infty(n):=\gamma_m(S)$. For fixed $n\ge3$ the exact tables are consistent
with an approach of order $1/N$, not $N^{-1/2}$ (Numerical observation~\ref{obs:cN}),
that is, with the unproved expansions
\[
	\gamma_1(N,n)=\gamma_1^\infty(n)-\frac{c_1(n)}{N}+o(N^{-1}),
	\qquad
	\gamma_2(N,n)=\gamma_2^\infty(n)-\frac{c_2(n)}{N}+o(N^{-1}).
\]
For $n\ge3$ we do not prove these, nor the existence of the coefficients; for $n=2$ they
hold on each parity class of $N$, with parity-dependent coefficients, and are proved
below. A symmetry heuristic explains why no $N^{-1/2}$ term
appears: the $N^{-1/2}$ term of the local expansion of the characteristic function of the edge vector is odd~\cite[Lemma~2.5]{banic_elezovic_p2}, so the corresponding Edgeworth term is an odd polynomial times an even Gaussian density,
while the smooth (Edgeworth) part of every moment of $d$ is an integral of a function that is even under the
reflection $\tilde D\mapsto-\tilde D$ (not the support parity of
Proposition~\ref{prop:support}), as for the mean and the
variance~\cite[Theorem~5.5]{banic_elezovic_p1} (for $d^c$, proportional to $d$ under the uniform null),~\cite[Lemma~6.2]{banic_elezovic_p2}
(see also~\cite[Lemma~5.4]{banic_elezovic_p3}); lattice corrections from the kink of
$|\cdot|$ are expected at order $1/N$. Turning this into a proof for the skewness and
the excess kurtosis would need a lattice Edgeworth expansion for the piecewise linear
functional $\sum|x_i|$, including the corrections at its kinks, with Fourier inversion
and lattice summation as in~\cite[Chapter~5]{bhattacharya_rao}; we do not give it.

\begin{observation}[The $1/N$ coefficients]\label{obs:cN}
On the grid of Section~\ref{ssec:grid}, the log--log slope of
$\gamma_m^\infty(n)-\gamma_m(N,n)$ over $100\le N\le500$ is between $-1.014$ and $-0.995$
for $m=1,2$ and $n\in\{3,5,10,20,50\}$. At $N=500$, $N(\gamma_1^\infty-\gamma_1)$ is
$0.880\sqrt n$ at $n=3$ and decreases to $0.830\sqrt n$ at $n=50$, i.e.\ the relative
skewness error is about $(0.50\text{--}0.54)\,n/N$; and $N(\gamma_2^\infty-\gamma_2)$ is
$3.39$ at $n=3$, $3.99$ at $n=4$, and between $4.06$ and $4.12$ for $5\le n\le50$. The
limit is reached only for $N\gg n$: at $n=50$, $\gamma_1^\infty=0.217104$, while the
exact skewness is $0.19749$, $0.20241$ and $0.20536$ at $N=300,400,500$, that is,
$9.04\%$, $6.77\%$ and $5.41\%$ below the limit.
\end{observation}

For $n=2$ the expansions hold, with coefficients that depend on the parity of $N$.
Here $d=|S_N|$ with $S_N=2X_1-N$, and the limit is the half-normal law, with
$\gamma_1^\infty=\sqrt2(4-\pi)/(\pi-2)^{3/2}=0.99527$ and
$\gamma_2^\infty=8(\pi-3)/(\pi-2)^2=0.86918$.

\begin{proposition*}[The case $n=2$]
On each parity class of $N$,
$\gamma_m(N,2)=\gamma_m^\infty-c_m/N+O(N^{-2})$ for $m=1,2$, with
\[
	c_1^{\rm even}=\frac{\sqrt2\,\pi(8-\pi)}{4(\pi-2)^{5/2}},\quad
	c_1^{\rm odd}=\frac{\sqrt2\,\pi(5\pi-16)}{4(\pi-2)^{5/2}},\quad
	c_2^{\rm even}=\frac{2\pi(\pi^2-8)}{(\pi-2)^3},\quad
	c_2^{\rm odd}=\frac{2\pi(4-\pi)^2}{(\pi-2)^3},
\]
for even and odd $N$.
\end{proposition*}

\begin{proof}
$S_N$ is a simple symmetric random walk. Let $\mu_N=\E|S_N|$, $p_N=\Pr(S_N=0)$,
$m=\lfloor N/2\rfloor$ and $e=N-2m$. Conditioning on $S_N$ gives
$\mu_{N+1}=\mu_N+p_N$ and $\E|S_{N+1}|^3=\E|S_N|^3+3\mu_N+p_N$. Since $p_N=0$ for odd
$N$ and $p_{2m}=\binom{2m}m4^{-m}$, induction on the first recurrence (with
$\binom{2m+2}{m+1}=\frac{2(2m+1)}{m+1}\binom{2m}m$) gives $\mu_N=N\binom{2m}m4^{-m}$, so
$\mu_N=Np_N$ for even $N$, and induction on the second then gives
$\E|S_N|^3=(2N-e)\mu_N$. Moreover $\E S_N^2=N$ and $\E S_N^4=3N^2-2N$. With $z=1/N$
and $a=\mu_N/\sqrt N$ the raw moments of $d/\sqrt N$ are $a$, $1$, $(2-ez)a$ and
$3-2z$, so that
\[
	\kappa_2=1-a^2,\qquad \kappa_3=2a^3-a-eza,\qquad
	\kappa_4=4a^2-6a^4-2z+4eza^2 .
\]
By Stirling's formula $\binom{2m}m4^{-m}=(\pi m)^{-1/2}\{1-1/(8m)+O(m^{-2})\}$, hence
$a=\sqrt{2/\pi}\,\{1+(2e-1)z/4+O(z^2)\}$ on each parity class. Since
$\kappa_2\to1-2/\pi>0$, substituting into $\gamma_1=\kappa_3\kappa_2^{-3/2}$ and
$\gamma_2=\kappa_4\kappa_2^{-2}$ and expanding to first order in $z$ gives the two
limits and the four coefficients.
\end{proof}

Extrapolations in $1/N$ of the exact cumulants at $N\le4001$ agree
with these values to at least seven digits. Since $c_1^{\rm odd}<0$,
for odd $N$ the skewness approaches its limit from above. For $n=3$ the fits indicate that the leading coefficients do not depend on the residue of $N$ modulo $3$: on the exact laws for
$970\le N\le1000$, least-squares fits of $\gamma_m^\infty(3)-\gamma_m(N,3)$ by
$a_m/N+b_m/N^2+e_m/N^3$, with the theoretical intercept fixed and made separately in
each residue class of $N$ modulo $3$ ($10$ or $11$ values of $N$), give
$a_1=c_1(3)=1.5297$ and $a_2=c_2(3)=3.3901$ in all three classes, while $b_m$ depends
on the class (classes $1$ and $2$ agree, class $0$ differs): in these fits the residue enters only at order $N^{-2}$. (With a free intercept and a quadratic in $1/N$ the fitted $a_1$ no
longer agree to four decimals across the classes, so fixing the theoretical limit
matters.) For $n\ge4$ we have not identified a residue
dependence of the leading coefficient.

The residue classes enter through the lattice of the edge vector. Since
$N=nX_1+\sum_{i=1}^{n-1}(n-i)D_i$, every attainable $(D_1,\ldots,D_{n-1})$ satisfies
$\sum_{i=1}^{n-1}(n-i)D_i\equiv N\pmod n$, so the edge vector lives on one coset,
determined by $N$ modulo $n$, of a sublattice of index $n$ in $\mathbb Z^{n-1}$; for
$n=2$ this is the parity $D_1\equiv N\pmod 2$. A finite-$N$ expansion therefore has to
track both the kinks of $\sum|x_i|$ and this coset.

Conversely, extrapolating the exact tables to $N\to\infty$ checks the cumulant
theory of Section~\ref{sec:cumulants}. Table~\ref{tab:extrap} compares the limits of
Theorem~\ref{thm:cumulants} with a least-squares fit $a+b/N+c/N^2$ to the exact values
at $N\ge100$. The agreement is about $10^{-6}$ for $n=5$ and $n=10$. At $n=3$ the fit
must be made within each residue class of $N$ modulo $3$ (three points each, so these fits are interpolations); the three
extrapolations then agree with the theory to $4\cdot10^{-7}$, whereas one fit over the
mixed grid is off by $3\cdot10^{-4}$. At $n=50$, where $N/n\le10$, the agreement is
about $2\cdot10^{-4}$.

\begin{table}[ht]
\centering\small
\caption{Limiting standardized cumulants: Theorem~\ref{thm:cumulants} against the
extrapolation $a+b/N+c/N^2$ of the exact values at $100\le N\le500$ (for $n=3$ the
mean of the fits made separately in each residue class of $N$ modulo $3$).}
\label{tab:extrap}
\begin{tabular}{rcccc}
\toprule
 & \multicolumn{2}{c}{$\gamma_1^\infty$} & \multicolumn{2}{c}{$\gamma_2^\infty$}\\
$n$ & theory & extrapolated & theory & extrapolated\\
\midrule
3  & 1.017616 & 1.017616 & 1.296227 & 1.296227\\
5  & 0.744786 & 0.744786 & 0.685721 & 0.685721\\
10 & 0.503113 & 0.503112 & 0.313509 & 0.313508\\
50 & 0.217104 & 0.216897 & 0.058372 & 0.058239\\
\bottomrule
\end{tabular}
\end{table}

\section{The gamma diagnostic}
	\label{sec:gamma}

For the normal-theory sum $D$, Kamat~\cite{kamat1953} used a Pearson Type~III (shifted
gamma) curve fitted to three moments for $n>8$, and Sathe and
Kamat~\cite{sathe_kamat1957} a power of a $\chi^2$ variable. For the finite multinomial
statistic $d$ the gamma law was introduced and recommended
in~\cite{banic_elezovic_ct}, where $29$ candidate families were fitted to the exact null
distributions by a discrete Cram\'er--von Mises criterion, and it is used for
$d_{\mathbf r}$ in~\cite{banic_elezovic_gct}. Here we analyse the canonical gamma law
matched to the exact mean and variance, and why and where it works. For a
nondegenerate law ($\sigma_N>0$; the only exception under the uniform null is $N=1$,
$n=2$) let $\mu_N=\E d$, $\sigma_N^2=\Var d$ (for $d_{\mathbf r}$ the same definitions
apply with $d_{\mathbf r}$ in place of $d$). The moment-matched gamma law has shape
$\nu_N=\mu_N^2/\sigma_N^2$ and scale $\theta_N=\sigma_N^2/\mu_N$, and its skewness
and excess kurtosis are
\begin{equation}\label{eq:gamma-skew-kurt}
	\gamma_{1,\Gamma}=\frac{2}{\sqrt{\nu_N}}=\frac{2\sigma_N}{\mu_N},
	\qquad
	\gamma_{2,\Gamma}=\frac{6}{\nu_N}=\frac{6\sigma_N^2}{\mu_N^2}.
\end{equation}
The first test of gamma behaviour beyond the two fitted moments is the pair of ratios
\begin{equation}\label{eq:gamma-cumulant-ratios}
	R_{3,N}=\frac{\gamma_{1,N}}{\gamma_{1,\Gamma}},
	\qquad
	R_{4,N}=\frac{\gamma_{2,N}}{\gamma_{2,\Gamma}},
\end{equation}
where $\gamma_{1,N}$ and $\gamma_{2,N}$ are the exact skewness and excess kurtosis. For a
gamma law both equal one. If they differ from one, the gamma law may still be a good
CDF approximation, but its accuracy is not explained by the higher moments; and
since the gamma excess kurtosis is always positive, a negative $\gamma_{2,N}$ rules out
an exact gamma law and agreement of the fourth cumulants, but does not by itself rule
out a close gamma CDF or accurate upper quantiles. Under the uniform null
Corollary~\ref{cor:asymptotics} gives the limits $R_3\to0.8370$ and $R_4\to0.5783$ as
$N\to\infty$ and then $n\to\infty$. A beta law on a fixed support $[0,L]$ ($L=2N$, the
range of $d$ for $n\ge3$; $L=N$ for $n=2$) also fixes its higher cumulants by its first
two moments. It exists, with positive shape parameters, exactly when $0<\mu_N<L$ and
$0<\sigma_N^2<\mu_N(L-\mu_N)$; under the uniform null this fails only for $N=1$, $n=2$
(degenerate) and for $N=n=2$, where $d\in\{0,2\}$ with equal probabilities is a
two-point law at the endpoints, a limit of beta laws. So neither law is an exact model,
and both have to be judged by their CDF and quantile errors.

\begin{table}[ht]
\centering\small
\caption{The moment-matched gamma threshold for the limit law $S$ at $n=50$, from a
Monte Carlo sample of $10^8$ draws of~\eqref{eq:S-msd}: standardized quantile shift
$\Delta_\alpha=(q^\Gamma_{1-\alpha}-q^S_{1-\alpha})/\sqrt{\Var S}$, realized level
$\Pr(S>q^\Gamma_{1-\alpha})$ and its relative error. Monte Carlo standard errors:
$0.002\%$, $0.001\%$ and $0.0003\%$ for the realized levels, and about $0.0002$,
$0.0004$ and $0.001$ for $\Delta_\alpha$ (Numerical observation~\ref{obs:limit-size}).}
\label{tab:limit-mc}
\begin{tabular}{cccc}
\toprule
nominal $\alpha$ & $\Delta_\alpha$ & realized $\alpha$ & relative error\\
\midrule
$5\%$   & $+0.011$ & $4.90\%$  & $-2\%$\\
$1\%$   & $+0.033$ & $0.928\%$ & $-7\%$\\
$0.1\%$ & $+0.071$ & $0.082\%$ & $-18\%$\\
\bottomrule
\end{tabular}
\end{table}

\begin{observation}[Effect at the level of the limit law]\label{obs:limit-size}
For the limit law at $n=50$ ($\E S=7.8193$, $\sqrt{\Var S}=1.0123$, gamma shape
$(\E S)^2/\Var S=59.67$; skewness $0.2171$ against the gamma's $0.2589$), a Monte Carlo
sample of $10^8$ draws of~\eqref{eq:S-msd} gives, for the gamma threshold at
nominal level $\alpha$, the standardized quantile shift
$\Delta_\alpha=(q^\Gamma_{1-\alpha}-q^S_{1-\alpha})/\sqrt{\Var S}$ and the realized
level listed in Table~\ref{tab:limit-mc}. At $n=50$ the gamma test is conservative even in the limit, by $7\%$ of the level at $1\%$ and by $18\%$ at $0.1\%$ (this is not so at every $n$: for $n=3$ at $5\%$ the limit law gives $1.013\alpha$). At $n=10$ the limit law gives $-2.7\%$, $-13.4\%$ and $-34.3\%$ (Monte Carlo,
$3\cdot10^9$ draws pooled over three independent runs, standard errors of these relative errors below $0.05$ percentage points), and the finite case
$N=200$ of Section~\ref{ssec:worked} gives $-2.6\%$, $-15.7\%$ and $-36.5\%$ (exact
table, $c=\lceil x\rceil$).
\end{observation}

\section{Evidence from exact tables and the tail approximation}
	\label{sec:tails}

Performance statements in this section are numerical observations on exact tables (or
on Monte Carlo simulation where stated) and are not proved; the approximation formulas
and the half-span rule are constructions, not theorems.

\subsection{The uniform grid}
	\label{ssec:grid}

For the uniform null we computed the exact law of $d$ on the grid
\[
	n\in\{2,3,4,5,6,7,8,9,10,12,20,30,50\},\qquad
	N\in\{10,20,\ldots,100,150,200,250,\ldots,500\},
\]
$13\times18=234$ cells, in exact integer arithmetic (exact laws for all
$2\le N,n\le500$ were computed in~\cite{banic_elezovic_ct}; the algorithm is described
in~\cite{banic_elezovic_p08}). The corrected quantile $q_p^{\Gamma+\mathrm{CF}}$
tabulated in Tables~\ref{tab:n10} and~\ref{tab:N200} is defined
by~\eqref{eq:gamma-CF-correction} in Section~\ref{ssec:worked}, which also defines the
half-span rule used later (Tables~\ref{tab:worked}--\ref{tab:calib}) to turn a
continuous approximate quantile into a critical value. As a
measure of the gamma CDF error we use the
continuity-corrected lattice discrepancy
\[
	\mathrm{KS}_{1/2}=\max_{k\in\operatorname{supp}d}\bigl|F_N(k)-G(k+\tfrac h2)\bigr| ,
\]
where $F_N$ is the exact CDF, $G$ the moment-matched gamma CDF, and $h$ the lattice step
of the statistic in the body of the law: $h=1$ for $d$ with $n\ge3$ (all values
reported below, apart from $n=2$), $h=2$ for $d$
with $n=2$ and for the even-valued $d^c$ with $n\ge3$ (Proposition~\ref{prop:support};
for $n=2$, $d^c=2|2X_1-N|$ has span $4$), and $h=H/L$ for $d_{\mathbf r}$ with $n\ge3$, the actual span of the exact support
(Section~\ref{sec:exact-finite}; for $n=2$ see~\cite{banic_elezovic_p08}); the
subscript of $\mathrm{KS}_{1/2}$ refers to the midpoint, i.e.\ the shift $h/2$. The
shift is half the lattice step,
not $\frac12$: for the even-valued $d^c$ the midpoint between the support points $k$
and $k+2$ is $k+1$, and the shift $\frac12$ would add a lattice error of the order of
a quarter of an atom, which would then dominate the discrepancy.
$\mathrm{KS}_{1/2}$ is taken over the support points only and is not
a Kolmogorov distance: the supremum over all $x$ is dominated by the lattice jumps and
is several times larger.

\begin{observation}[The regular subgrid]\label{obs:subgrid}
On the $129$ cells with $N\ge50$, $n\ge5$ and $N\ge3n$, $\mathrm{KS}_{1/2}$ lies between
$0.0036$ and $0.0098$ (median $0.0072$), and the skewness ratio $R_3$ lies between
$0.68$ and $0.85$ (median $0.816$): the exact laws show the skewness deficit predicted
by Corollary~\ref{cor:asymptotics}. In the sparse corner the picture changes: the
exact uniform law has negative excess kurtosis in $15$ cells, all with $N/n\le4/3$
(for example $\gamma_2=-0.202$ at $N=10$, $n=20$, and $-0.022$ at $N=n=50$), the regime
of Remark~\ref{rem:sparse-axis}.
\end{observation}

\begin{table}[ht]
\centering\small
\caption{Uniform null, $n=10$: gamma CDF error, cumulant ratios and quantiles. Here $q_p$
is the exact $p$-quantile (the exact critical value is $q_p+1$), $q^\Gamma_p$ the gamma
quantile and $q^{\Gamma+\mathrm{CF}}_p$ the corrected quantile~\eqref{eq:gamma-CF-correction}.
The corrected quantile uses the exact finite-$N$ skewness and excess kurtosis of the
same exact law, so it is a diagnostic (an oracle input); the version with the limiting
cumulants is calibrated in Table~\ref{tab:calib}.}
\label{tab:n10}
\begin{tabular}{rcccccccc}
\toprule
$N$ & $\mathrm{KS}_{1/2}$ & $R_3$ & $R_4$ & $q_{.95}$ & $q^{\Gamma}_{.95}$ & $q_{.99}$ & $q^{\Gamma}_{.99}$ & $q^{\Gamma+\mathrm{CF}}_{.99}$ \\
\midrule
50 & 0.0090 & 0.753 & 0.433 & 34 & 34.54 & 40 & 40.97 & 40.16 \\
100 & 0.0074 & 0.799 & 0.510 & 49 & 49.07 & 57 & 58.19 & 57.23 \\
150 & 0.0069 & 0.814 & 0.537 & 60 & 60.19 & 70 & 71.37 & 70.27 \\
200 & 0.0066 & 0.822 & 0.550 & 69 & 69.56 & 81 & 82.47 & 81.24 \\
250 & 0.0065 & 0.826 & 0.558 & 78 & 77.80 & 91 & 92.24 & 90.90 \\
300 & 0.0064 & 0.829 & 0.563 & 85 & 85.25 & 100 & 101.07 & 99.62 \\
350 & 0.0063 & 0.832 & 0.567 & 92 & 92.10 & 108 & 109.19 & 107.64 \\
400 & 0.0063 & 0.833 & 0.569 & 98 & 98.47 & 115 & 116.75 & 115.11 \\
450 & 0.0062 & 0.835 & 0.572 & 104 & 104.46 & 122 & 123.84 & 122.11 \\
500 & 0.0062 & 0.836 & 0.573 & 110 & 110.12 & 129 & 130.55 & 128.74 \\
\bottomrule
\end{tabular}
\end{table}

For fixed $n=10$ (Table~\ref{tab:n10}) the gamma CDF error is already small at $N=50$
and decreases slowly with $N$, and $R_3$, $R_4$ increase towards their limits
$0.8448$ and $0.5893$ (Table~\ref{tab:gamma-n}). The $95\%$ quantile is well tracked by
the gamma law. At $99\%$ the gamma quantile is already about one to two lattice
steps too high, and the Cornish--Fisher correction removes most of this. At $n=50$ the same
trend is slower: from $N=300$ to $N=500$ ($N/n=6$ to $10$) $\mathrm{KS}_{1/2}$ decreases
from $0.0041$ to $0.0036$, while $R_3$ and $R_4$ increase from $0.760$ and $0.442$ to
$0.792$ and $0.497$, still below the limits $0.8385$ and $0.5805$.

\begin{table}[ht]
\centering\small
\caption{Uniform null, $N=200$: as Table~\ref{tab:n10} (corrected quantile with the
exact finite-$N$ cumulants), across $n$, at $p=.95$ and $p=.999$.}
\label{tab:N200}
\begin{tabular}{rcccccccc}
\toprule
$n$ & $\mathrm{KS}_{1/2}$ & $R_3$ & $R_4$ & $q_{.95}$ & $q^{\Gamma}_{.95}$ & $q_{.999}$ & $q^{\Gamma}_{.999}$ & $q^{\Gamma+\mathrm{CF}}_{.999}$ \\
\midrule
3 & 0.0186 & 0.854 & 0.610 & 40 & 39.19 & 65 & 70.58 & 65.37 \\
4 & 0.0087 & 0.848 & 0.589 & 46 & 46.23 & 74 & 77.72 & 73.05 \\
5 & 0.0081 & 0.842 & 0.581 & 51 & 51.61 & 79 & 82.71 & 78.57 \\
6 & 0.0078 & 0.837 & 0.574 & 56 & 56.08 & 83 & 86.71 & 82.96 \\
7 & 0.0074 & 0.832 & 0.567 & 60 & 59.96 & 87 & 90.15 & 86.69 \\
8 & 0.0071 & 0.829 & 0.561 & 63 & 63.44 & 90 & 93.22 & 89.99 \\
9 & 0.0069 & 0.825 & 0.555 & 66 & 66.62 & 93 & 96.03 & 92.99 \\
10 & 0.0066 & 0.822 & 0.550 & 69 & 69.56 & 96 & 98.64 & 95.75 \\
12 & 0.0063 & 0.816 & 0.539 & 75 & 74.89 & 101 & 103.40 & 100.75 \\
20 & 0.0055 & 0.794 & 0.502 & 92 & 91.92 & 117 & 118.93 & 116.79 \\
30 & 0.0051 & 0.769 & 0.458 & 108 & 108.21 & 132 & 134.15 & 132.28 \\
50 & 0.0048 & 0.721 & 0.374 & 133 & 133.18 & 156 & 157.94 & 156.27 \\
\bottomrule
\end{tabular}
\end{table}

\begin{figure}[htbp]
\centering
\includegraphics[width=0.78\linewidth]{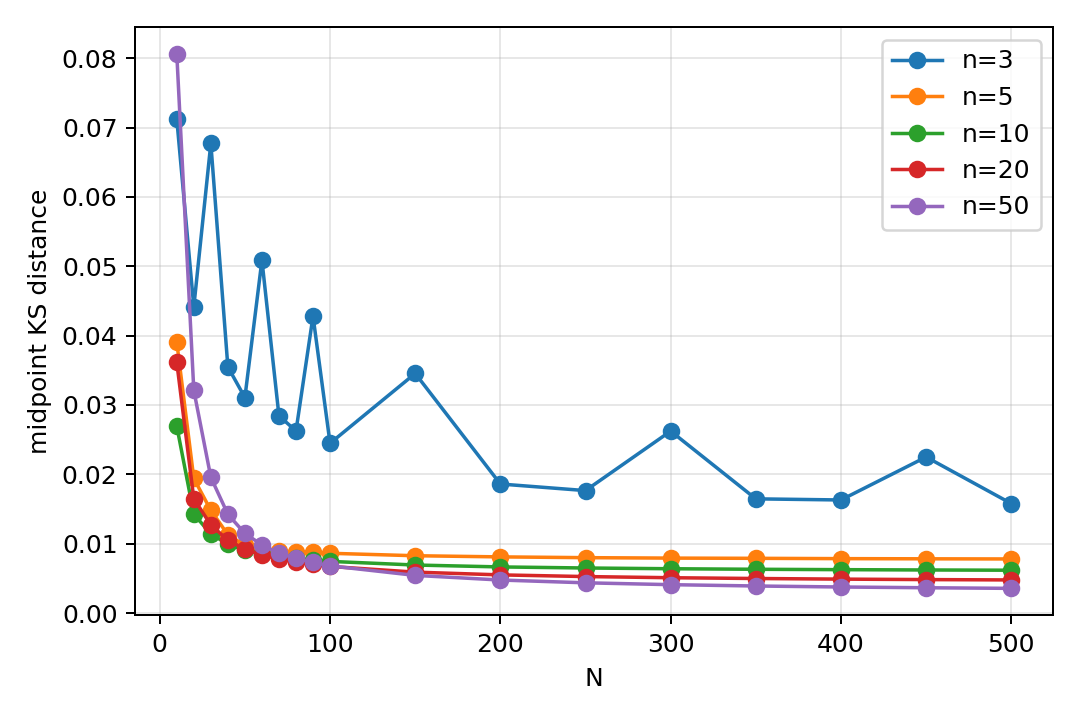}
\caption{The continuity-corrected lattice discrepancy $\mathrm{KS}_{1/2}$ between the exact
uniform DTV law and the moment-matched gamma law, for $n=3,5,10,20,50$ (the $n=3$ curve
stays oscillatory).}
\label{fig:KS-gamma}
\end{figure}

\begin{figure}[htbp]
\centering
\includegraphics[width=0.78\linewidth]{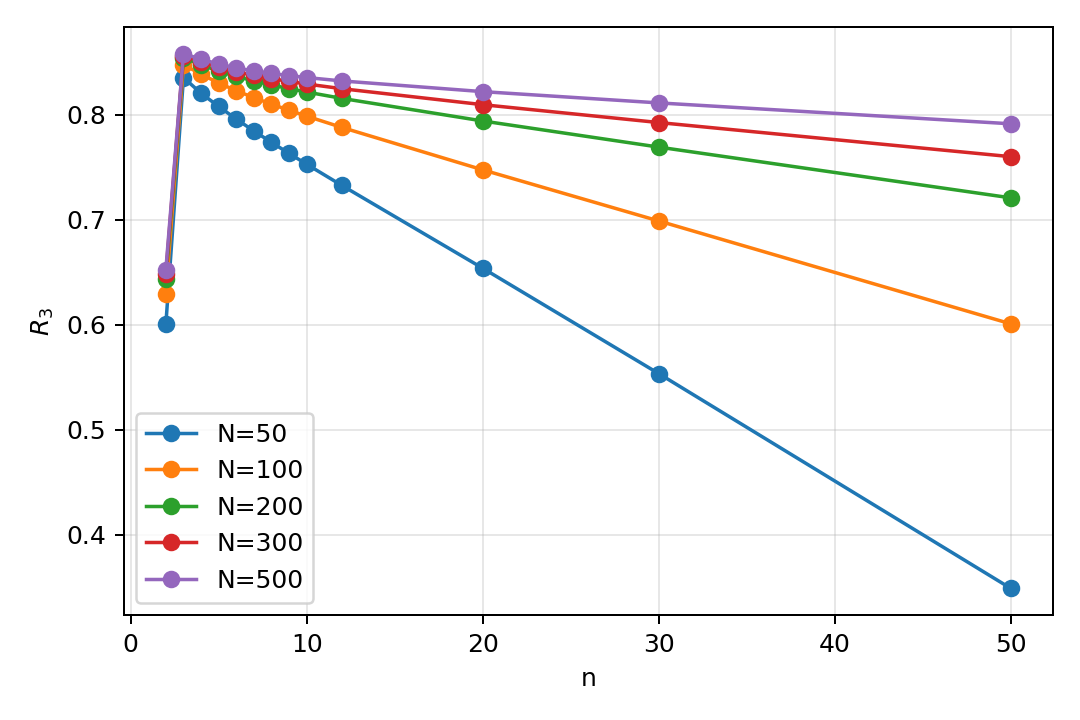}
\caption{Exact skewness divided by the gamma skewness on the grid, for
$N=50,100,200,300,500$.
The deficit $R_3<1$ is the finite-$N$ counterpart of Corollary~\ref{cor:asymptotics};
for fixed $N$ the fall at large $n$ also moves the cells towards the sparse regime.}
\label{fig:R3}
\end{figure}

\begin{figure}[htbp]
\centering
\includegraphics[width=0.78\linewidth]{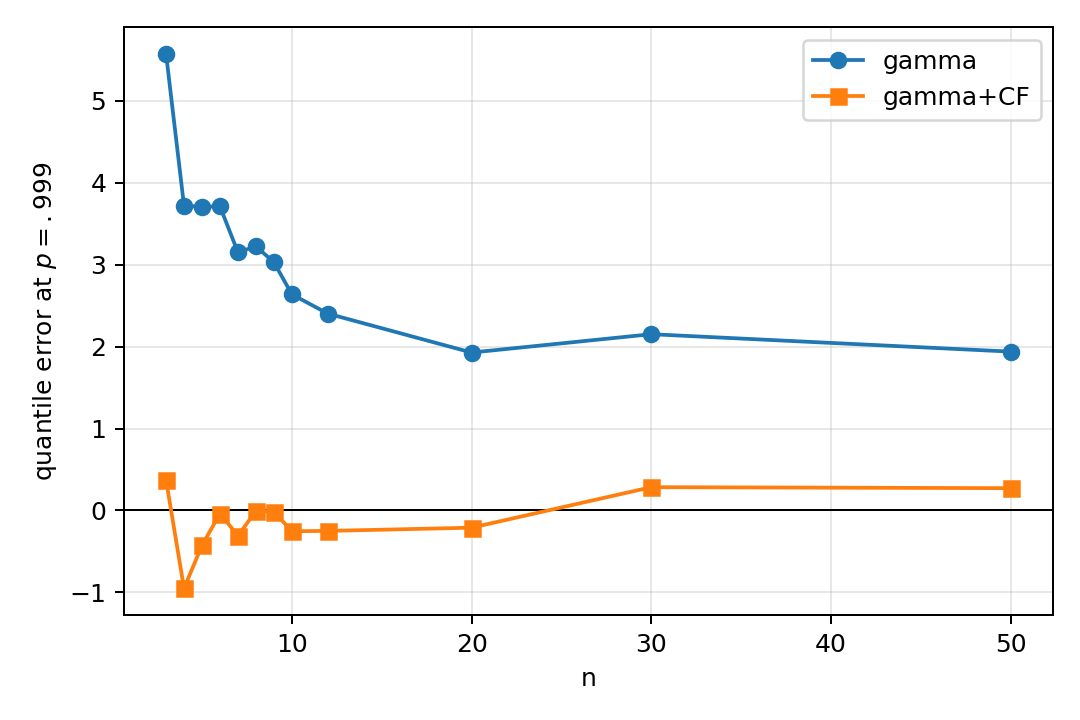}
\caption{Quantile error at $p=.999$, $N=200$, in lattice steps: the gamma quantile and
the corrected quantile~\eqref{eq:gamma-CF-correction} minus the exact quantile.}
\label{fig:q999-error}
\end{figure}

\begin{figure}[htbp]
\centering
\includegraphics[width=0.78\linewidth]{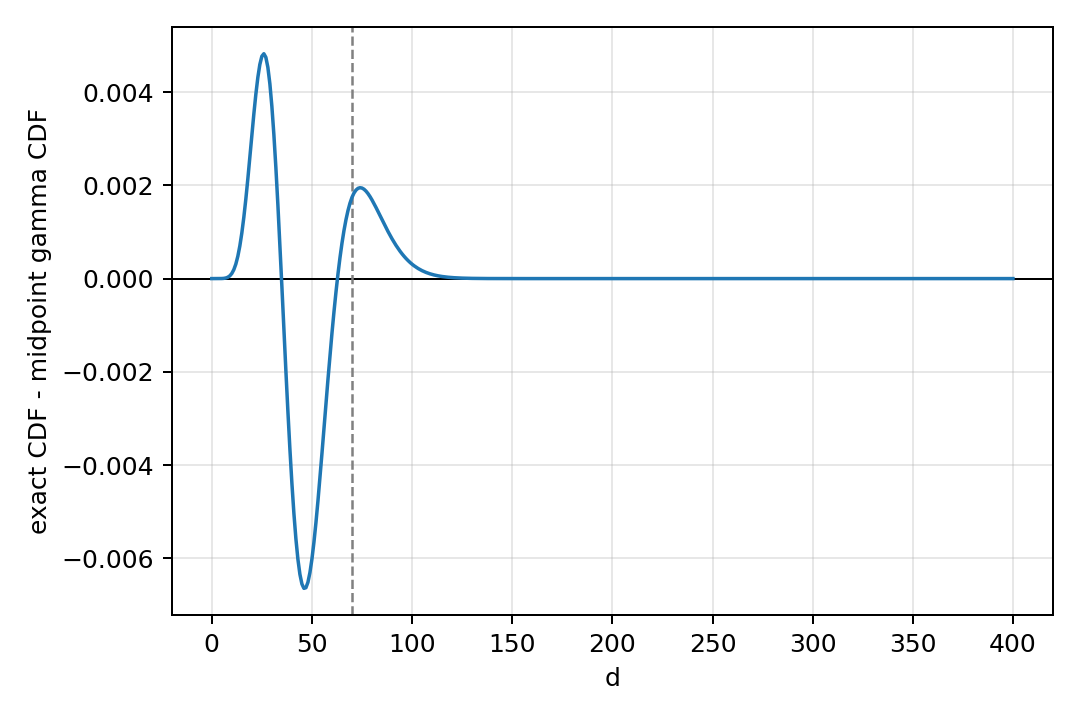}
\caption{CDF residual for $N=200$, $n=10$: exact CDF minus the continuity-corrected
gamma CDF, $F_N(k)-G(k+\frac12)$, at the support points (joined by a line). The
vertical line marks the exact $5\%$ critical value $d=70$.}
\label{fig:cdf-residual}
\end{figure}

For fixed $N=200$ (Table~\ref{tab:N200}) the largest gamma CDF error among $n\ge3$ is at
$n=3$, and for $n\ge5$ $\mathrm{KS}_{1/2}<0.01$. The ratios $R_3$ and $R_4$ stay below
one, and at $p=.999$ the gamma quantile is too high by about two to six steps, while the
corrected quantile is within one step. Figures~\ref{fig:KS-gamma}--\ref{fig:q999-error}
show the grid data graphically, and Figure~\ref{fig:cdf-residual} the CDF residual of
the worked example.

\FloatBarrier

\subsection{Non-uniform profiles}
	\label{ssec:profiles}

Table~\ref{tab:profiles} shows exact standardized cumulants of the raw statistic for
four alternating integer profiles $1,2,1,\ldots$ (the null $\mathbf r$ proportional
to the listed weights), and two uniform cyclic laws for comparison.

\begin{table}[ht]
\centering\small
\caption{Exact cumulant diagnostics of the raw statistic for alternating profiles
(all $\Delta_i\ne0$), and of the uniform cyclic statistic. The last column is
$z=\min_i\sqrt N|\Delta_i|/\sqrt{\Var G^{(\mathbf r)}_i}$, the distance of the edge means
from $0$ in edge standard deviations.}
\label{tab:profiles}
\begin{tabular}{llrrrrrrr}
\toprule
statistic & profile & $N$ & $\gamma_1$ & $\gamma_2$ & $\gamma_{1,\Gamma}$ & $R_3$ & $R_4$ & $z$\\
\midrule
linear & $1,2,1$ & 400 & $0.0000$ & $-0.0050$ & $0.3000$ & $0.0000$ & $-0.0370$ & 6.0\\
linear & $1,2,1,2,1$ & 300 & $-0.0220$ & $-0.0084$ & $0.3391$ & $-0.0650$ & $-0.0485$ & 3.9\\
linear & $1,2,1,2,1,2,1$ & 250 & $-0.0178$ & $-0.0344$ & $0.3664$ & $-0.0487$ & $-0.1708$ & 2.9\\
cyclic & $1,2,1,2,1,2$ & 150 & $0.0010$ & $-0.0999$ & $0.4541$ & $0.0022$ & $-0.3229$ & 2.4\\
cyclic, uniform & $n=10$ & 50 & $0.4248$ & $0.1993$ & $0.5757$ & $0.7379$ & $0.4009$ & --\\
cyclic, uniform & $n=20$ & 50 & $0.2620$ & $0.0590$ & $0.4083$ & $0.6418$ & $0.2359$ & --\\
\bottomrule
\end{tabular}
\end{table}

For the alternating profiles the skewness is far below the gamma value and the excess
kurtosis is negative, which no gamma law has. The linearization of
Remark~\ref{rem:linear-cumulants} explains the first two rows: for $1,2,1$ it gives
$\gamma_1=0$ and $\gamma_2=-2/N=-0.0050$ exactly for the linearized statistic, and up
to exponentially small terms for $d$, and for $1,2,1,2,1$ it gives
$\gamma_1=-0.0229$ and $\gamma_2=-0.0056$ against the exact $-0.0220$ and $-0.0084$.
For the two longer profiles the edge means are only $2.4$--$2.9$ edge standard
deviations from zero at these $N$, so sign locking is incomplete and some edges still
fold. Since every $\Delta_i\ne0$, the fixed-$n$ limit is the linear Gaussian regime
(Theorem~\ref{thm:regimes} for the linear rows; for the cyclic row the same
sign-locking argument applies to the $n$ cyclic edges); these rows are still pre-asymptotic, in the transition from
folding to linearization, and their negative excess kurtosis ($-0.034$ and $-0.100$, against the linear values $-0.0065$
and $-0.0100$) comes mainly from the folding. The uniform cyclic laws keep positive
skewness and excess kurtosis, both below the gamma values. So raw $d$ under a
non-uniform null and the centered or uniform statistic need different approximations.
This is the sign locking of~\cite{banic_elezovic_gct}, which is why the comb test under
a non-uniform null is defined there through the centered $d_{\mathbf r}$; for
$d_{\mathbf r}$ the limit law is folded for every profile (Theorem~\ref{thm:limit}(b)),
and its cumulants are those of Section~\ref{ssec:profile}.

\subsection{A worked example and the Cornish--Fisher correction}
	\label{ssec:worked}

Take the uniform null with $N=200$, $n=10$. The exact table gives
$\mu=45.3024491486$, $\sigma^2=182.3249192962$, $\gamma_1=0.489866$ and
$\gamma_2=0.292961$; the moment-matched gamma law has shape $\nu=11.2563433829$,
$\theta=4.0246150644$, $\gamma_{1,\Gamma}=0.596117$ and $\gamma_{2,\Gamma}=0.533033$. The
gamma law has larger skewness and larger excess kurtosis; consistently with this, in
this example it puts too much mass into the far right tail (Table~\ref{tab:worked}; see
also Proposition~\ref{prop:tail}).

Let $z_p$ be the standard normal $p$-quantile. The second-order Cornish--Fisher
expansion~\cite{cornish_fisher1938,fisher_cornish1960}, the formal inversion of the
Edgeworth series, is
\begin{equation}\label{eq:CF-correction}
	C(z;\gamma_1,\gamma_2)
	=
	z+\frac{\gamma_1}{6}(z^2-1)
	+\frac{\gamma_2}{24}(z^3-3z)
	-\frac{\gamma_1^2}{36}(2z^3-5z),
\end{equation}
and the normal Cornish--Fisher quantile is $q_p^{\mathrm{CF}}=\mu+\sigma
C(z_p;\gamma_1,\gamma_2)$. Keeping the gamma law as the baseline and correcting only
the difference between the exact and the gamma cumulants gives
\begin{equation}\label{eq:gamma-CF-correction}
	q_p^{\Gamma+\mathrm{CF}}
	=
	q_p^\Gamma
	+
	\sigma\bigl[
	C(z_p;\gamma_1,\gamma_2)
	-
	C(z_p;\gamma_{1,\Gamma},\gamma_{2,\Gamma})
	\bigr],
\end{equation}
a gamma-anchored difference of two normal Cornish--Fisher expansions: since the target
and the matched gamma law have the same $\mu$ and $\sigma$, subtracting their expansions
removes the common terms, and the gamma quantile replaces the normal one as the
baseline (for expansions about a general base see~\cite{hill_davis1968}). When the exact cumulants equal the gamma cumulants, the
bracket vanishes. We use~\eqref{eq:gamma-CF-correction} as an empirical correction, not as a theorem.

\paragraph{The critical value.}
For a lattice law of span one and a continuous approximation $G$ matched at the
midpoints, $G(k+\frac12)\approx F_N(k)$, an approximate $p$-quantile $x$ is to be compared
with the exact quantile $q$: a perfect approximation has
$x\in(q-\frac12,q+\frac12]$ and $|x-q|$ spread over $[0,\frac12]$, with median about
$\frac14$ (the benchmark if $x-q$ is spread uniformly over the cell). The exact critical value, the smallest $c$ with
$\Pr(d\ge c)\le1-p$, is $q+1$. We therefore take
\[
	c=\Bigl\lceil x+\tfrac12\Bigr\rceil .
\]
With $c=\lceil x\rceil$ an approximation whose error is spread uniformly over the cell is anti-conservative about half of the time.

The same argument for a lattice of step $h$, matched by $G(k+\frac h2)\approx F_N(k)$,
gives the \emph{half-span rule}: $c$ is the smallest support point with $c\ge x+\frac h2$ ($c=+\infty$, i.e.\ no rejection, if there is none). A perfect approximation then has $x\in(q-\frac h2,q+\frac h2]$, and
the rule returns the next support point above $q$, the exact critical value, provided
the support is a full progression of step $h$ near $q$. Where the support has holes,
such as $d=1$ for $n\ge4$ or the band above $N$ for $n=3$, the natural placement is the
midpoint between consecutive support points, and the half-span shift is a practical
surrogate for it~\cite{banic_elezovic_p08}; the shift itself uses the global span $h$,
not the local spacing $3$ above $N$ for $n=3$, so there the rule is a convention, and
Table~\ref{tab:across} reports its performance. The shift is
$\frac12$ for $d$ ($n\ge3$), $1$ for $d$ with $n=2$ and for the even-valued $d^c$
($n\ge3$; $2$ for $d^c$ with $n=2$), and $H/(2L)$ for $d_{\mathbf r}$ with $n\ge3$,
whose actual span $H/L$ is read from the exact support and equals $s/L$ for $N\ge N_0$
(Section~\ref{sec:exact-finite} and~\cite{banic_elezovic_p08}). With the naive shift
$1/(2L)$ the test for $d_{\mathbf r}$ can be anti-conservative: for $\mathbf r=(1,2,1)/4$,
$N=200$ (step $1$), the normal Cornish--Fisher test with the exact cumulants of
$d_{\mathbf r}$ at $N=200$ has, at $1\%$, size $1.31\alpha$ with the
shift $\frac18$ and $0.87\alpha$ with the shift $\frac12$~\cite{banic_elezovic_p08}
(we recomputed this case from the exact law of Proposition~\ref{prop:n3}(a)). The
correct shift is necessary but not sufficient at small $n$: in the same example the
$5\%$ test has size $1.12\alpha$.

\begin{table}[ht]
\centering\small
\caption{Uniform null, $N=200$, $n=10$. Exact quantile $q$ and the deviations $x-q$ of
the gamma, the corrected~\eqref{eq:gamma-CF-correction} and the normal Cornish--Fisher
quantiles (the last two with the exact finite-$N$ cumulants), and of the corrected quantile with the limiting
$\gamma_1(S),\gamma_2(S)$ of Theorem~\ref{thm:cumulants} and the exact mean and variance (last column); in brackets the realized size $\Pr(d\ge c)/\alpha$ of the test with
$c=\lceil x+\frac12\rceil$.}
\label{tab:worked}
\begin{tabular}{rrrrrr}
\toprule
$\alpha$ & $q$ & gamma & gamma$+$CF & normal CF & gamma$+$CF, limiting\\
\midrule
$5\%$      & 69  & $+0.56\ [0.86]$  & $+0.24\ [0.97]$ & $+0.25\ [0.97]$ & $+0.28\ [0.97]$\\
$1\%$      & 81  & $+1.47\ [0.84]$  & $+0.24\ [0.98]$ & $+0.28\ [0.98]$ & $+0.37\ [0.98]$\\
$0.1\%$    & 96  & $+2.64\ [0.53]$  & $-0.25\ [0.90]$ & $-0.13\ [0.90]$ & $+0.02\ [0.90]$\\
$10^{-4}$  & 109 & $+4.33\ [0.38]$  & $-0.53\ [1.01]$ & $-0.28\ [0.84]$ & $-0.09\ [0.84]$\\
$10^{-5}$  & 120 & $+7.13\ [0.20]$  & $+0.06\ [0.92]$ & $+0.47\ [0.92]$ & $+0.66\ [0.74]$\\
$10^{-6}$  & 131 & $+9.31\ [0.10]$  & $-0.17\ [0.84]$ & $+0.45\ [0.84]$ & $+0.61\ [0.67]$\\
$10^{-7}$  & 141 & $+12.04\ [0.04]$ & $-0.03\ [0.82]$ & $+0.84\ [0.64]$ & $+0.94\ [0.64]$\\
\bottomrule
\end{tabular}
\end{table}

At $5\%$ the gamma quantile is $69.555$; the rule $\lceil x\rceil$ gives the exact
critical value $70$, with $\Pr(d\ge70)=0.04870186$, but $\lceil x+\frac12\rceil=71$ is
one step conservative. With $\lceil x\rceil$ the plain gamma test has realized size
$0.97\alpha$, $0.84\alpha$ and $0.63\alpha$ at $\alpha=5\%,1\%,0.1\%$. The gamma
quantile is too high by $1.5$ steps at $1\%$ and by $12$ steps at $10^{-7}$
(Table~\ref{tab:worked}), qualitatively consistent, over this range, with the
lighter-than-gamma (Gaussian) tail of the limit law (Proposition~\ref{prop:tail}). The corrected quantile (with the exact cumulants) stays within about half a step of the exact quantile down to $\alpha=10^{-7}$,
and with $c=\lceil x+\frac12\rceil$ it gives the exact critical value at every level of
Table~\ref{tab:worked} except $10^{-4}$ ($x=108.47$, $c=109=q$). The normal
Cornish--Fisher quantile is as good down to $10^{-6}$ and drifts upwards below. With the
limiting cumulants the corrected quantile is as good down to $10^{-4}$ but is $0.66$, $0.61$
and $0.94$ steps high at $10^{-5}$, $10^{-6}$ and $10^{-7}$ (last column of
Table~\ref{tab:worked}); on the $84$ grid cells with $N\ge200$ and $n\ge3$ its median
absolute error is $0.35$, $0.54$, $1.00$ and $1.28$ steps at $\alpha=10^{-4}$, $10^{-5}$,
$10^{-6}$, $10^{-7}$. So ``within about half a step down to $10^{-7}$'' needs the exact cumulants.

The Cornish--Fisher \emph{quantile} and the Edgeworth \emph{CDF} are not equivalent in
the deep tail. As a diagnostic we substitute the exact finite-$N$ cumulants
into~\eqref{eq:edgeworth-limit-law} and add a continuity correction; no error transfer
from the limit law is claimed. This gives $\Pr(d\ge110)\approx3.9\cdot10^{-5}$,
$\Pr(d\ge120)\approx1.8\cdot10^{-6}$ and $\Pr(d\ge130)\approx4.3\cdot10^{-8}$, against
the exact $8.4\cdot10^{-5}$, $1.13\cdot10^{-5}$ and $1.31\cdot10^{-6}$ (the gamma values
are $1.9\cdot10^{-4}$, $3.6\cdot10^{-5}$ and $6.7\cdot10^{-6}$). In this example, below about $10^{-3}$ the Edgeworth CDF is strongly anti-conservative, while the gamma-anchored quantile~\eqref{eq:gamma-CF-correction} remains accurate.

\subsection{Across the grid}
	\label{ssec:across}

Table~\ref{tab:across} repeats the comparison on all cells of the grid with $N\ge40$
and $N\ge3n$, and at $p=.999$ only $N\ge100$, so that the tail quantile is resolved. Errors
are in lattice steps (span $2$ for $n=2$); the critical value is given by the half-span rule of Section~\ref{ssec:worked} (the smallest support point $c\ge x+\frac h2$, i.e.\ $c=\lceil x+\frac12\rceil$
for $n\ge3$ and shift $1$ for $n=2$); it counts as exact if it defines the same test as
the exact critical value, and as
anti-conservative if $\Pr(d\ge c)>1-p$. All corrected quantiles in
Tables~\ref{tab:n10}--\ref{tab:across} use the exact finite-$N$ skewness and excess
kurtosis;
this is a diagnostic, since in the present computations these are read off the exact law (the first four moments alone also follow from a moment-only recurrence~\cite{banic_elezovic_p08}). The rule with the limiting
values is reported separately in Table~\ref{tab:calib}.

\begin{table}[ht]
\centering\small
\caption{Gamma and gamma$+$CF quantiles on the grid (uniform null, exact finite-$N$
cumulants): median signed and absolute quantile error and maximal absolute error in
lattice steps, and the critical value $c$ = the smallest support point
$\ge x+\frac h2$ (half-span rule).}
\label{tab:across}
\setlength{\tabcolsep}{4pt}
\begin{tabular}{rr|cc|cc|cc|cc|cc}
\toprule
 & & \multicolumn{2}{c|}{median err} & \multicolumn{2}{c|}{median $|$err$|$} & \multicolumn{2}{c|}{max $|$err$|$}
   & \multicolumn{2}{c|}{\% exact c.v.} & \multicolumn{2}{c}{\% anti-cons.}\\
$p$ & cells & gamma & $+$CF & gamma & $+$CF & gamma & $+$CF & gamma & $+$CF & gamma & $+$CF\\
\midrule
$0.95$  & 181 & $+0.19$ & $-0.10$ & 0.27 & 0.29 & 1.21 & 1.61 & 74 & 90 & 2 & 9\\
$0.99$  & 181 & $+1.16$ & $-0.15$ & 1.16 & 0.25 & 2.82 & 1.93 & 7  & 83 & 0 & 15\\
$0.999$ & 116 & $+3.43$ & $-0.27$ & 3.43 & 0.36 & 7.82 & 1.66 & 0  & 67 & 0 & 32\\
\bottomrule
\end{tabular}
\end{table}

At $5\%$ the gamma quantile is within about a quarter of a step in median, but it
misses the exact critical value in a quarter of the cells, mostly on the conservative
side; the correction fixes most of these. At $1\%$ the gamma quantile is about one step
high and gives the exact critical value in only $7\%$ of the cells, and at $0.1\%$ it is
$3.4$ steps high (median) and never exact. The corrected quantile has median absolute error
$0.25$--$0.36$ steps at all three levels ($0.29$, $0.26$ and $0.32$ with the limiting
cumulants), close to this benchmark $\frac14$, but its
signed median error is negative, so it errs towards the anti-conservative side. For fixed $n$ the gamma error in steps grows like $\sqrt N$ wherever its standardized version tends to a nonzero constant, since $\sigma$ grows like $\sqrt N$ (at $p=.99$ and $.999$ this constant is positive in the cases we checked; at $p=.95$ it can be negative, $-0.010$ for $n=3$); on the grid this is visible for
$n\le10$ (at $p=.999$ and $n=5$ it grows from $2.5$ steps at $N=100$ to $5.8$ at
$N=500$), while for $n=50$ the range $N/n\le10$ is too short. At $p=.999$ the normal
Cornish--Fisher quantile has a smaller median absolute error
than~\eqref{eq:gamma-CF-correction} ($0.22$ against $0.36$ steps on the $116$ cells
with $N\ge100$, $N\ge3n$) and is markedly better calibrated as a test (exact critical
value in $91\%$ against $67\%$ of the cells, anti-conservative in $7\%$ against $32\%$).
With the limiting cumulants the same holds for $n\ge5$: at $0.1\%$ the normal
Cornish--Fisher critical value is exact in $81\%$ of the $89$ cells and never
anti-conservative (largest size $0.976\alpha$), against $76\%$, $9\%$ and $1.041\alpha$
for~\eqref{eq:gamma-CF-correction} (Table~\ref{tab:calib}). Deeper in the tail the gamma
baseline is better: on the $70$ cells with $N\ge200$ and $n\ge5$, with the limiting
cumulants and the half-span rule, \eqref{eq:gamma-CF-correction} gives the exact critical
value in $70\%$, $53\%$ and $27\%$ of the cells at $\alpha=10^{-4},10^{-5},10^{-6}$
(anti-conservative only at $10^{-4}$, in $4\%$ of the cells, with largest size
$1.029\alpha$), and the normal Cornish--Fisher quantile in $39\%$, $4\%$ and $0\%$, always
conservative (largest sizes $0.935\alpha$, $0.893\alpha$ and $0.816\alpha$).

\begin{table}[ht]
\centering\small
\caption{Calibration of the critical-value rules on the grid of Table~\ref{tab:across}
(uniform null): percentage of cells with the exact critical value, percentage of
anti-conservative cells, and the largest realized size $\Pr(d\ge c)/\alpha$. Half: the
smallest support point $c\ge x+\frac h2$; full: $c\ge x+h$. ``Exact'' and ``limiting''
refer to the values of $\gamma_1,\gamma_2$ used in the correction (the limiting ones
from Theorem~\ref{thm:cumulants}, with the exact mean and variance).}
\label{tab:calib}
\setlength{\tabcolsep}{4pt}
\begin{tabular}{l|ccc|ccc|ccc}
\toprule
 & \multicolumn{3}{c|}{$\alpha=5\%$} & \multicolumn{3}{c|}{$\alpha=1\%$}
 & \multicolumn{3}{c}{$\alpha=0.1\%$}\\
rule & exact & anti & max & exact & anti & max & exact & anti & max\\
\midrule
\multicolumn{10}{l}{all cells ($181$, $181$, $116$)}\\
gamma, half & 74 & 2 & 1.142 & 7 & 0 & 0.974 & 0 & 0 & 0.734\\
$\Gamma$+CF exact, half & 90 & 9 & 1.142 & 83 & 15 & 1.193 & 67 & 32 & 1.472\\
$\Gamma$+CF exact, full & 59 & 1 & 1.031 & 56 & 5 & 1.177 & 73 & 7 & 1.213\\
$\Gamma$+CF limiting, half & 89 & 7 & 1.142 & 81 & 12 & 1.193 & 66 & 22 & 1.472\\
$\Gamma$+CF limiting, full & 55 & 1 & 1.031 & 46 & 5 & 1.177 & 47 & 6 & 1.128\\
normal CF exact, half & 92 & 8 & 1.142 & 87 & 12 & 1.193 & 91 & 7 & 1.128\\
normal CF limiting, half & 91 & 6 & 1.142 & 81 & 10 & 1.193 & 77 & 6 & 1.128\\
\midrule
\multicolumn{10}{l}{$n\ge5$ ($136$, $136$, $89$ cells)}\\
gamma, half & 71 & 0 & 0.974 & 7 & 0 & 0.907 & 0 & 0 & 0.734\\
$\Gamma$+CF exact, half & 97 & 3 & 1.002 & 94 & 6 & 1.030 & 80 & 20 & 1.060\\
$\Gamma$+CF exact, full & 56 & 0 & 0.974 & 53 & 0 & 0.983 & 76 & 0 & 0.976\\
$\Gamma$+CF limiting, half & 93 & 3 & 1.002 & 90 & 3 & 1.030 & 76 & 9 & 1.041\\
$\Gamma$+CF limiting, full & 51 & 0 & 0.965 & 42 & 0 & 0.983 & 40 & 0 & 0.962\\
normal CF exact, half & 97 & 3 & 1.002 & 96 & 4 & 1.009 & 99 & 1 & 1.002\\
normal CF limiting, half & 93 & 2 & 1.002 & 88 & 1 & 1.005 & 81 & 0 & 0.976\\
\bottomrule
\end{tabular}
\end{table}

\paragraph{Practical rule.}
The corrected quantile~\eqref{eq:gamma-CF-correction} is an approximation of a
quantile, not a test with guaranteed level. With the exact cumulants and the half-span
rule it gives the exact critical value in $90\%$, $83\%$ and $67\%$ of the grid cells at
$\alpha=5\%,1\%,0.1\%$, but it is anti-conservative in $9\%$, $15\%$ and $32\%$ of them,
with largest sizes $1.14\alpha$, $1.19\alpha$ and $1.47\alpha$ (Table~\ref{tab:calib});
its signed median error is negative, $-0.10$ to $-0.27$ steps. The failures sit at small
$n$: the worst cell is $n=2$, $N=500$, $\alpha=0.1\%$ ($x=70.67$, $c=72$,
$\Pr(d\ge72)=1.47\cdot10^{-3}$, against the exact critical value $76$), and at $n=3$
the size reaches $1.19\alpha$ ($N=300$, $\alpha=1\%$). For $n\ge5$ the
largest size is $1.06\alpha$, and for $n\ge10$ it is $1.02\alpha$, with the exact
critical value in at least $97\%$ of the cells. With the limiting
$\gamma_1(S),\gamma_2(S)$ of Theorem~\ref{thm:cumulants} and the exact mean and variance
of~\cite{banic_elezovic_p1,banic_elezovic_p2} the rule needs no exact table; on the grid
it behaves similarly (anti-conservative in $7\%$, $12\%$ and $22\%$ of the cells, largest
size $1.47\alpha$ at $n=2$; none for $n\ge10$ at $1\%$ and $0.1\%$).

We therefore recommend three tiers. In the second and the third tier the
quantiles are computed with the limiting $\gamma_1(S),\gamma_2(S)$ of
Theorem~\ref{thm:cumulants} and the exact mean and variance; the rows of Table~\ref{tab:calib} with the exact finite-$N$ cumulants, extracted here from the exact laws, are diagnostics. For $n\le4$ use the exact law: for $n=2$ that of $|2X_1-N|$ with
$X_1\sim\mathrm{Bin}(N,\frac12)$, for $n=3$ it needs only binomial distribution
functions (Proposition~\ref{prop:n3}(a)), and for $n=4$ it is computed
in~\cite{banic_elezovic_p08}. For $n\ge5$, when an approximate critical value is
acceptable, use the half-span rule with a quantile that depends on the level: the normal
Cornish--Fisher quantile at $\alpha\ge10^{-3}$ and the corrected
quantile~\eqref{eq:gamma-CF-correction} at $\alpha\le10^{-4}$ (levels in between were not
tabulated). On the grid, for $n\ge5$, the normal Cornish--Fisher quantile gave the exact
critical value in $93\%$, $88\%$ and $81\%$ of the cells at $\alpha=5\%,1\%,0.1\%$,
anti-conservative in $2\%$, $1\%$ and $0\%$ of them, with largest size $1.005\alpha$ (the
corrected quantile: $93\%$, $90\%$ and $76\%$, anti-conservative in $3\%$, $3\%$ and $9\%$,
largest size $1.04\alpha$); at $10^{-4}$, $10^{-5}$ and $10^{-6}$ the corrected quantile is
the better one (see above).

A more conservative empirical
option is the full-span shift, the smallest support point $c\ge x+h$ (often, but not always, the support point after the half-span one; row ``$\Gamma$+CF limiting, full''): at
the three tabulated levels on the grid it was never anti-conservative for $n\ge5$
(largest size $0.98\alpha$), at
the price of a median size of $0.84$--$0.88\alpha$ and the exact critical value in only
$40$--$51\%$ of the cells (with the exact cumulants $0.86$--$0.91\alpha$ and
$53$--$76\%$). This is an observation on a finite grid at three levels, not a level
guarantee: at the untabulated level $\alpha=0.000777$ and $(N,n)=(500,5)$ the same rule
gives $c=128$ with exact size $1.001\alpha$, whereas the exact critical value is $129$.
Over the whole range $10^{-4}\le\alpha\le0.05$, evaluated at the jump points of $c$ on
the cells with $n\ge5$ and $N\ge3n$, the largest size with the limiting cumulants is
$1.002\alpha$ for the full-span rule and $1.055\alpha$ for the half-span rule, both at
$(N,n)=(500,5)$. For
$n\le4$ the variant is not safe (up to $1.18\alpha$; $1.21\alpha$ with the exact
cumulants). When the size must not exceed $\alpha$, the approximation alone does not suffice: evaluate
the exact upper tail at the proposed support point and move upwards until it is at
most $\alpha$, or use the exact critical value directly. A Monte Carlo test~\cite{hope1968} also has level at most $\alpha$, at the cost of simulation.

These statements concern the
uniform null: under a non-uniform null, on the $182$ tables
of~\cite{banic_elezovic_p08} (design below), the
full-span variant (with the exact cumulants) stayed below $\alpha$ only for
$n\ge7$ (up to $1.15\alpha$ for $n=5,6$ and $1.30\alpha$ for $n\le4$). Returning to the
uniform grid of Table~\ref{tab:calib}, the plain gamma law was conservative in every cell at the two tabulated levels $1\%$ and $0.1\%$ (not at every smaller level: at $(N,n)=(60,3)$ and $\alpha=0.0086$ its size is $1.014\alpha$), but at $0.1\%$ its median size is only
$0.55\alpha$; for $\alpha=5\%$, with the exact mean and variance and
$c=\lceil x+\frac12\rceil$, it gives the exact critical value in $74\%$ of the cells and
is conservative in most of the others (with $c=\lceil x\rceil$ it is exact in $70\%$ and
anti-conservative in $30\%$). All these are observations on $N\le500$, $N\ge3n$; when the
exact critical value is needed, or for $p$-values far below $10^{-3}$, use the exact law,
which is available for $2\le N,n\le500$~\cite{banic_elezovic_ct,banic_elezovic_p08}.
For a lattice of step $h$ the shift $\frac12$ is replaced by $\frac h2$: by $1$ for $d$
with $n=2$ and for the cyclic statistic with $n\ge3$ (by $2$ for $d^c$ with $n=2$), and
by $H/(2L)$ for $d_{\mathbf r}$ with $n\ge3$, where $H/L$ is the actual span of the exact
support ($H=s$ for $N\ge N_0$, with $s$ given explicitly in Section~\ref{sec:exact-finite}
and~\cite[Theorem~3.1]{banic_elezovic_p08}; below $N_0$ the span is read off the exact
support, Section~\ref{sec:exact-finite}); the same shift is used in $\mathrm{KS}_{1/2}$.
Under a non-uniform null the rule is less reliable: the companion
paper~\cite{banic_elezovic_p08} reports, on $182$ exact tables of $d_{\mathbf r}$, the
exact critical value in only $39$--$53\%$ of the cells at $\alpha=5\%,1\%,0.1\%$, with
sizes up to $1.49\alpha$ at $n=3$ and at most $1.04\alpha$ for $n\ge7$ on the $22$
profiles of its calibration set. That set consists of all exact tables with $N\ge50$
for the profiles $\mathbf r=\mathbf q/L$, $L=\sum_iq_i$, with $\mathbf q$ equal to
$(1,2,1)$, $(1,2,3)$, $(1,3,1)$, $(1,4,1)$; $(1,2,1,2)$, $(1,2,3,2)$, $(1,2,3,4)$,
$(1,4,2,3)$; $(1,2,1,2,1)$, $(1,2,3,2,1)$, $(1,2,3,4,5)$, $(1,3,1,3,1)$, $(1,3,2,4,1)$,
$(2,4,1,3,1)$; $(1,2,1,2,1,2)$, $(3,1,2,4,1,2)$, $(3,1,4,1,5,2)$; $(1,2,1,2,1,2,1)$;
$(1,2)$ repeated $5$ and $6$ times, $(1,2,3)$ repeated $4$ times and $(1,2,3,2)$ repeated
$3$ times. Here $N\in\{50,60,\dots,100,150,200,250\}$ for $n\le5$, the same set up to
$200$ for $n=6,7$, and up to $100$ for $n=10,12$ ($14\cdot9+4\cdot8+4\cdot6=182$
tables, all with $N\ge3n$ and $\min_iNr_i\ge2.08$); the shift uses the actual span $H/L$
of each exact support. For degenerate, spiky profiles with small expected
counts ($\min_iNr_i\lesssim1$) the rule can fail badly, up to size $1$, so there the
exact law, or at least the exact tail at the proposed critical value, is
required~\cite{banic_elezovic_p08}. For $n=3$ the exact law needs only binomial
distribution functions (Proposition~\ref{prop:n3}(a)).

\section{Discussion}
	\label{sec:discussion}

\emph{Proved for fixed $n$.} The support and the parity of the exact law
(Proposition~\ref{prop:support}); the folded limit of $d_{\mathbf r}/\sqrt N$ for every
null with all $r_i>0$, with convergence of all moments (Theorem~\ref{thm:limit}); and,
for the raw statistic under a null with all $r_i>0$, the limit of
$(d-N\sum_{i\in I_1}|\Delta_i|)/\sqrt N$, after removal of the deterministic drift: folded
under the uniform null, linear Gaussian when every $\Delta_i\ne0$, and mixed otherwise
(Theorem~\ref{thm:regimes}); for $n=2$, the $1/N$ expansions of the skewness and the
excess kurtosis on each parity class of $N$ (Section~\ref{sec:finite-N}).

\emph{Proved for the limit laws.} Under the uniform null, the cumulants through order
four in elementary closed form; their moments are Kamat's~\cite{kamat1953}, and the ingredient we add is the identification of his numerically evaluated integral as $\pi^2/30$, with a short proof of the closed form of the four-edge moment (Lemma~\ref{lem:E1234}; the closed form is due to Finch~\cite{finch2015}); further the
exactness for the cyclic statistic and the Gaussian extreme tail. Under a null with all
$r_i>0$, $\kappa_3(Q_{\mathbf r})$ for every such fixed profile, and the leading
large-$n$ terms of the skewness and the excess kurtosis for profiles generated by a
fixed smooth positive $f$; $\kappa_4(Q_{\mathbf r})$ at fixed $n\ge5$ is only
algorithmic. Profiles with structural zeros are not covered.

\emph{Observed, not proved.} The $O(1/N)$ approach of the skewness and the excess
kurtosis for $n\ge3$ and the $n=3$ residue fits (Section~\ref{sec:finite-N}), the shape statements of Numerical observations~\ref{obs:shape} and~\ref{obs:subgrid}, the $n=3$ range of Numerical observation~\ref{obs:n3-range}, the limit-law sizes of Numerical observation~\ref{obs:limit-size}, and the calibration results of Section~\ref{sec:tails}.

\emph{Practical.} The corrected quantile~\eqref{eq:gamma-CF-correction} is a quantile
approximation, accurate on the tested uniform grid ($N\le500$, $N\ge3n$; median absolute error $0.26$--$0.32$ of a lattice step at $\alpha=5\%,1\%,0.1\%$ with the limiting cumulants, $0.25$--$0.36$ with the exact finite-$N$ cumulants). As a critical value for $n\ge5$ it is the better choice at $\alpha\le10^{-4}$, while at $\alpha\ge10^{-3}$ the plain normal Cornish--Fisher quantile with the limiting cumulants is better calibrated (at $0.1\%$ never anti-conservative on the grid, against $9\%$ of the cells). Neither has a size guarantee, also in the full-span
variant, and both are weaker for small $n$ and for non-uniform nulls. Certified level control needs the exact law~\cite{banic_elezovic_p08}, or at least its upper tail at the proposed critical value, or a Monte Carlo test; the exact law gives the sharp critical value and remains the reference.

\section{Open problems}
	\label{sec:open}

\begin{enumerate}
\item \textbf{Higher cumulants.} The connected-cluster constants $a_{m,L}$ of order
$m\ge5$ (the structure of $\kappa_m(S)$ and $\kappa_m(S_c)$ is given after
Proposition~\ref{prop:cyclic}). For each $m$ they are computable from multivariate
normal orthant probabilities~\cite{kanrobotti2017}, and their sign ingredients
$\E\prod_i\sgn Z_i$ are rational, being counts of ascent patterns of random
permutations (Remark~\ref{rem:status}). Open are their elementary evaluation and
whether, for every $m$ and $L\le m$, they lie in the $\mathbb Q$-algebra generated by
$\pi^{\pm1/2}$, square roots of positive integers and arcsines of rational numbers, as
for $m\le4$. The transfer operator of Remark~\ref{rem:transfer} gives the per-edge cumulants $\ell_m$
numerically.

\item \textbf{The $1/N$ terms.} A proof, for $n\ge3$, of the $O(1/N)$ approach of the
skewness and the excess kurtosis (Section~\ref{sec:finite-N}), on residue classes of $N$
where needed, and the coefficients in closed form; for $n=3$ the leading coefficient appears not to depend on the residue,
and for $n\ge4$ this is not known. Possible tools are the local expansion of the
multinomial law~\cite{ouimet2021}, summed over the sign cones of $\sum|x_i|$ and over
the attainable coset of Section~\ref{sec:finite-N}, the lattice expansions for sums in
convex sets~\cite{yarnold1972} and continuity-corrected Edgeworth
series~\cite{kolassa_mccullagh1990}; the case $n=2$, proved in Section~\ref{sec:finite-N} from exact binomial moments, with
coefficients that depend on the parity of $N$, is a test case for such a method.

\item \textbf{The mixed regime.} For the limit~\eqref{eq:mixed-limit} with
$I_0,I_1\ne\varnothing$, write it as $A+L$ with $A$ folded and $L$ Gaussian; by symmetry
$\kappa_3=\kappa_3(A)+3\kappa(A,L,L)$ and $\kappa_4=\kappa_4(A)+6\kappa(A,A,L,L)$. The
reduction to absolute and signed Gaussian moments gives the two mixed terms
algorithmically; their elementary forms (in the sense of Section~\ref{ssec:profile})
in the edge covariance, their signs and their
profile dependence are open.

\item \textbf{The fourth cumulant off the uniform at fixed $n$.} The cumulants of
$Q_{\mathbf r}$ under a positive general null are settled up to
$\kappa_3$ for every profile (Theorem~\ref{thm:kappa3-profile}), up to $\kappa_4$ for
$n\le4$, and, for smooth positive profiles, in the leading large-$n$ terms of
$\kappa_3,\kappa_4$ (Theorem~\ref{thm:profile-law}). What remains is a compact form of
$\kappa_4(Q_{\mathbf r})$ at fixed $n\ge5$: it is computable from Nabeya's four-variate
formula~\cite{nabeya1961} (see also~\cite{kanrobotti2017}) as an $O(n^4)$ sum involving
quadrivariate orthant probabilities, and whether it has an elementary form in the sense
of Section~\ref{ssec:profile}, as under the uniform null, is open; and the infimum and supremum of
$\gamma_1(Q_{\mathbf r})$ over profiles with all $r_i>0$ for $n\ge4$ (for $n=3$ the numerical values of Numerical observation~\ref{obs:n3-range} are not certified).

\item \textbf{The ray regime.} Under the uniform null, along $N/n\to\rho\in(0,\infty)$
(Remark~\ref{rem:sparse-axis}), the following are not treated here: the
central limit theorem and the joint grand-canonical cumulants (through the Perron root
of a transfer kernel), the conditioned third and
fourth cumulants per cell, the first Edgeworth and Cornish--Fisher terms, a local limit
theorem (with span $2$ for $d^c$, and with two parity classes of unequal limiting mass
$\frac12(1\pm e^{-4\rho})$ for $d$), and sharp large deviations. The negative excess
kurtosis of the sparse cells (Numerical observation~\ref{obs:subgrid}) motivates
testing whether the ray-limit fourth cumulant changes sign.

\item \textbf{The deep tail.} For fixed $n$ and the uniform null, $d/N=V(\mathbf q)
=\sum_i|q_{i+1}-q_i|$ with $\mathbf q$ the empirical frequencies, and
$V(\mathbf q)=\max_{\varepsilon}a_\varepsilon^{\!\top}\mathbf q$ over
$\varepsilon\in\{\pm1\}^{n-1}$, where $a_{\varepsilon,j}=\varepsilon_{j-1}-\varepsilon_j$
(with $\varepsilon_0=\varepsilon_n=0$). By Sanov's theorem and the contraction
principle~\cite{csiszar1984}, for $0<x<\max V$ the logarithmic rate of $\Pr(d\ge Nx)$ is
\[
	I_n(x)=\min_{\varepsilon}\,\sup_{t\ge0}\Bigl\{tx-\log\frac1n\sum_{j=1}^ne^{t a_{\varepsilon,j}}\Bigr\},
\]
a minimum over finitely many sign cones of scalar Legendre transforms. A cone with
$x<\max_ja_{\varepsilon,j}$ contributes a finite value, attained at the tilted law
$q_j\propto e^{t a_{\varepsilon,j}}$ with a finite $t>0$; a cone with
$\max_ja_{\varepsilon,j}<x$ is infeasible and contributes $+\infty$; and at
$x=\max_ja_{\varepsilon,j}$ the value $\log(n/k)$, with $k$ the number of maximal
$a_{\varepsilon,j}$, is reached only as $t\to\infty$. (For $n=3$ and $x=\frac32$ three of
the four cones are infeasible; only $a_\varepsilon=(-1,2,-1)$ remains.) Since $\frac1n\sum_ja_{\varepsilon,j}^2$
is maximal, equal to $\sigma_*^2=4-6/n$, for the alternating $\varepsilon$,
$I_n(x)\sim x^2/(2\sigma_*^2)$ as $x\downarrow0$, which is consistent with
Proposition~\ref{prop:tail}. For a single cone the lattice prefactor is
classical~\cite{bahadur_rao1960}, and conditional saddlepoint approximations for
statistics of multinomial frequencies, as independent counts conditioned on their
total, are in~\cite{skovgaard1987,gatto_jammalamadaka1999}; see
also~\cite{daniels1954,lugannani_rice1980}. Open are the identification of the
minimising cones as $x$ varies and their switching points, the prefactor where several
cones tie, a lattice approximation with proved relative error for the finite-$N$ tail
beyond $10^{-5}$, and its uniform matching, through moderate deviations $d=t\sqrt N$,
with the Gaussian extreme tail of Proposition~\ref{prop:tail}.
\end{enumerate}

\medskip\noindent\textbf{Acknowledgements.} The authors used AI assistants in the preparation
of this paper: Anthropic's Claude Opus~5.5 for additional literature searches, symbolic and numerical
verification, and editorial support, and OpenAI's Codex Sol~6.1 for a critical review of an
earlier version and for assistance with the subsequent revision. All mathematical results
and all changes to the text were checked independently by the authors, who take full
responsibility for the content.

\medskip\noindent\textbf{Code and data.} The exact distributions were computed in exact
integer arithmetic. Scripts verify the constants of
Table~\ref{tab:constants} and regenerate Tables~\ref{tab:gamma-n}--\ref{tab:calib} and
the Monte Carlo of Numerical observation~\ref{obs:limit-size}.
Programs and data are available from the authors on request, and a common archive for
the series is being prepared. Formulas, exact quantiles and software for practical use are
the subject of the companion paper~\cite{banic_elezovic_p11}.

\end{document}